\documentclass[11pt,a4paper]{article}
\pdfoutput=1
\usepackage{jheppub}
\usepackage[utf8]{inputenc}
\usepackage{epstopdf}
\usepackage{graphicx}
\usepackage{epsfig}
\usepackage{dcolumn}  
\usepackage{bm}    
\usepackage{caption}
\usepackage{subcaption}
\usepackage{amssymb} 
\usepackage{amsmath,bm}
\usepackage{amsfonts}  
\usepackage{amsmath}  
\usepackage{slashed}  
\usepackage{enumitem}
\usepackage[mathscr]{euscript}
\usepackage{tabu}
\usepackage{epsfig}
\makeatletter
\g@addto@macro\bfseries{\boldmath}
\makeatother

\def\Re{R\'{e}nyi }
\def\l1{{{1-loop}}}

\def\n1{\Bigg|_{n=1}}

\def\n{{(n)}}

\usepackage[T1]{fontenc} 
\usepackage{tikz}
\usepackage{amsmath,amssymb}
\usepackage{relsize}
\usepackage{latexsym}
\usepackage{leftidx}
\usepackage{xcolor}
\usepackage{diagbox}
\usepackage[T1]{fontenc}
\usepackage{array}
\usepackage{makecell}
\usepackage{csquotes}
\usepackage{tikz}
\usepackage{enumitem}
\usepackage{setspace}
\usepackage{multirow}
\usepackage{amsmath,amssymb}
\usepackage{relsize}
\usepackage{latexsym}
\usepackage{leftidx}
\usepackage{xcolor}
\usepackage{csquotes}
\usepackage{tikz}
\usepackage{enumitem}
\usetikzlibrary{decorations.markings}
\usetikzlibrary{decorations.pathmorphing}
\usetikzlibrary{decorations.markings}
\usetikzlibrary{decorations.pathmorphing}
\usepackage{pifont}
\usepackage{hyperref}

\def\rh{\rho^{\psi_1|\psi_2}}
\def\ra{\rho^{\psi_1|\psi_2}_A}

\usepackage{bookmark}

\title{\boldmath Pseudo Entropy in $U(1)$ gauge theory}

 \title{\textbf{\textsf{ Pseudo Entropy in $U(1)$ gauge theory
}}}
  \author{Jyotirmoy Mukherjee}
\affiliation{\vspace{.1cm} Centre for High Energy Physics, \\ Indian Institute of Science,\\
C. V. Raman Avenue, Bangalore 560012, India.}
\emailAdd{ jyotirmoym@iisc.ac.in}

\abstract{We study the properties of pseudo entropy, a new generalization of entanglement entropy, in free Maxwell field theory in $d = 4$ dimension. We prepare excited states by the different components of the field strengths located at different Euclidean times acting on the vacuum. We compute the difference between the pseudo \Re entropy and the \Re entropy of the ground state and observe that the difference changes significantly near the boundary of the subsystems and vanishes far away from the boundary. Near the boundary of the subsystems, the difference between pseudo \Re entropy and \Re entropy of the ground state depends on the ratio of the two Euclidean times where the operators are kept. To begin with, we develop the method to evaluate pseudo entropy  of conformal scalar field  in $d=4$ dimension. We prepare two states by two operators with fixed conformal weight acting on the vacuum and observe that the difference between pseudo \Re entropy and ground state \Re entropy changes only near the boundary of the subsystems. We also show that a suitable analytical continuation of pseudo \Re entropy leads to the evaluation of real-time evolution of \Re entropy during quenches.  }

\begin{document} 
\maketitle
\flushbottom

\section{Introduction}
\label{sec:intro}
Entanglement entropy is a useful quantity in a quantum system to characterize the degrees of freedom present in the system. For conformal field theory in $2$-dimension, the universal contribution to the entanglement entropy of the ground state is proportional to the central charge \cite{Cardy:1988tk,Calabrese:2004eu,Calabrese:2009qy} or the degrees of freedom and the similar statements hold true in higher dimensions. Moreover, the holographic derivation of entanglement entropy \cite{Ryu:2006ef,Hubeny:2007xt} gives us a deeper insight into the gravity emerging from the quantum entanglement.

To define the entanglement entropy of a quantum system, one subdivides the Hilbert space into two Hilbert spaces and integrates out the degrees of freedom in one of the Hilbert spaces to define the reduced density matrix. Finally one obtains entanglement entropy as the Von-Neumann entropy of the reduced density matrix.
\begin{align}
    S_{A}=-\rm{Tr}_{A}\left(\rho_A\log\rho_A\right).
\end{align}
Here $\rho_{A}$ is the reduced density matrix associated with the sub-region A. Generally one computes the entanglement entropy of the ground state or vacuum of a quantum system by evaluating the Von-Neumann entropy of the reduced density matrix of the ground state. 

Recently a new generalisation of entanglement entropy known as the pseudo-entropy has been introduced in \cite{Nakata:2020luh,Mollabashi:2020yie,Mollabashi:2021xsd} which is a Von-Neumann entropy of the transition matrix $\rho^{\psi_1|\psi_2}$. The transition matrix is constructed from initial state $|\psi_1\rangle$ and a final state $|\psi_2\rangle$ of a quantum system, where $|\psi_1\rangle$ and $|\psi_2\rangle$ are not orthogonal to each other.
\begin{align}
   \rho^{\psi_1|\psi_2}=\frac{|\psi_1\rangle\langle\psi_2|}{\langle \psi_1|\psi_2\rangle}, \qquad\qquad\qquad\qquad\langle \psi_1|\psi_2\rangle \neq 0.
\end{align}
Now one subdivides the Hilbert spaces into two Hilbert spaces and traces out the degrees of freedom in one of the sub-spaces to define the reduced transition matrix
\begin{align}\label{red}
     \rho^{\psi_1|\psi_2}_A=\rm{Tr}_{B} \rho^{\psi_1|\psi_2}.
\end{align}
 Therefore, the pseudo-entropy is defined as the Von Neumann entropy of the reduced transition matrix
 \begin{align}
     S_A(\rho^{\psi_1|\psi_2}_A)=-\rm{Tr}\left( \rho^{\psi_1|\psi_2}_A\log  \rho^{\psi_1|\psi_2}_A\right).
 \end{align}
 When initial and final states are the same, pseudo-entropy reduces to entanglement entropy.
 Note that, the reduced transition matrix is non-Hermitian in general and pseudo-entropy can be complex-valued. But this quantity is useful in the post-selection process where a initial state results into a final state and one is interested in measuring the weak value \cite{PhysRevLett.60.1351,RevModPhys.86.307} of an observable $\langle \mathcal{O}\rangle=\rm{Tr}(\mathcal{O} \rho^{\psi_1|\psi_2})$ .
  
    We evaluate pseudo \Re entropy for different fields to understand its general properties. The main motivation of this paper is to study the properties of the pseudo-entropy in gauge theory, in particular free Maxwell theory in $4$-dimension. However, there are subtleties in defining the entanglement entropy of the ground state of a gauge theory because the degrees of freedom are non-local. But it has been understood well for the free Maxwell field in $4$-dimension \cite{Donnelly:2014fua,Casini:2015dsg,Casini:2019nmu,Ghosh:2015iwa,Soni:2016ogt,David:2020mls} and in linearized gravity \cite{Benedetti:2019uej,David:2020mls,David:2022jfd}. In free Maxwell theory, we prepare two excited states by different components of the field strengths acting on the vacuum. Therefore the excited states remain  gauge invariant and the pseudo-entropy of the two states becomes well defined. At a constant time slice, we subdivide the region by a planar boundary and study pseudo-entropy as a function of the distance from the boundary of the subsystems. We evaluate the difference between pseudo \Re entropy and \Re entropy of the ground state and observe that the difference is non-zero near the boundary of the subsystems and it vanishes  far away from the boundary. The difference between pseudo \Re entropy and the \Re entropy of the ground state near the boundary depends on the ratio of the Euclidean times where two operators are kept. This indicates that the pseudo \Re entropy and \Re entropy of the ground state are the same when two operators are far away from the boundary of the subsystems.
 
 The paper is organized as follows. In section \eqref{2}, we define pseudo-entropy in conformal field theory using the replica trick. In section\eqref{3}, we begin with revisiting the computation of pseudo-entropy in $d=2$ conformal scalar field theory. Moreover, we take a slightly different approach which is to vary the positions of the operators instead of the center of the subsystems which was done in \cite{Nakata:2020luh}. This approach results in the same conclusion since we have a translational invariance along the spatial direction. Then we move on to evaluating pseudo-entropy in $d=4$ dimension. We prepare two excited states by two operators which act on the vacuum and  study pseudo-entropy as a function of the distance from the boundary. In section \eqref{5}, we study the properties of pseudo-entropy in free Maxwell theory in $d=4$ dimension. We first create two different states by the same components of the field strengths acting on the vacuum located at two different Euclidean times. Similarly, we use two different field tes to prepare different states. In both cases, we study the behavior of pseudo-entropy as a distance from the boundary of the subsystems. We also study its real-time behavior of it by Wick rotating the Euclidean time to Minkowski time and observe that the difference of pseudo \Re entropy and the  \Re entropy of the ground state saturates to a constant value $\log 2$ after a large time.
 
 \section{Pseudo-entropy in conformal field theory}\label{2}
 Given two non-orthogonal states $|\psi_1\rangle$ and $|\psi_2\rangle$, one defines the pseudo-\Re entropy in the following way
 \begin{align}\label{def1}
    S^{(n)}_A(\ra)&=\frac{1}{1-n}\log \left(\rm{Tr} (\ra)^{n}\right), \qquad\qquad n\geq 2
 \end{align}
where $n$ is the \Re parameter and pseudo-entropy can be obtained by taking $n\rightarrow 1$ limit in the equation \eqref{def1}. The reduced transition matrix is defined in \eqref{red}. One can evaluate $ S_A^{(n)}(\rh)$ in quantum field theory using the replica trick method which is explained in details in \cite{Nakata:2020luh}. The inner product of the states is evaluated using the path integral approach on a manifold with proper boundary conditions imposed on the states. We denote the manifold corresponding to the inner product of the states $\langle \psi_1|\psi_2\rangle$ as $\Sigma_1$ and $\rm{Tr} (\ra)^{n}$ by $\Sigma_n$. Then, $n$th pseudo-\Re entropy can be expressed as \cite{Nakata:2020luh}
 \begin{align}
      S^{(n)}_A(\ra)&=\frac{1}{1-n}\log\frac{Z_{\Sigma_n}}{(Z_{\Sigma_1})^n},
 \end{align}
 where $Z_{\Sigma_n}$ corresponds to the path integral over the $n$-sheeted manifold.
 We are interested in evaluating the pseudo-entropy in conformal field theory. We create two states by two operators acting on vacuum at two different points.
 \begin{align}
     |\psi_1\rangle=\mathcal{N}_1\mathcal{O}_1(\textbf{x}_1,t_1)|0\rangle, \quad\quad     |\psi_2\rangle=\mathcal{N}_2\mathcal{O}_2(\textbf{x}_2,t_2)|0\rangle,
 \end{align}
 where $\mathcal{N}_1$ and $\mathcal{N}_2$ are the normalization constants. Therefore the reduced transition matrix becomes
 \begin{align}
 \ra&=\mathcal{N} \rm{Tr}_B\left(\mathcal{O}_1(\textbf{x}_1,t_1)|0\rangle\langle 0|\mathcal{O}^{\dagger}_2(\textbf{x}_2,t_2)\right)
 \end{align}
Here $\mathcal{N}$ is the overall constant to ensure the unit normalization of the reduced transition matrix.
 
We would like to ask how pseudo-entropy varies from the ground state of a conformal field theory. Therefore, we compute the difference between the pseudo\Re entropy and the \Re entropy of the ground state
 \begin{align}
     \Delta S_A^{(n)}&= S^{(n)}_A(\rho^{\psi_1|\psi_2}_A)-S^{(n)}_A(\rho_A^{(0)}),
 \end{align}
 where $\rho_A^{(0)}$ is the reduced density matrix of the vacuum ,i.e, $\rho_A^{(0)}=\rm{Tr}_B|0\rangle\langle 0|$. $\rm{Tr} (\ra)^{n}$ can be evaluated by performing path integral over $n$-sheeted manifold with two operators $\mathcal{O}_1$ and $\mathcal{O}_2$ inserted at each sheet but in different points. Finally the difference can be written in the following way \cite{Nakata:2020luh}
 \begin{align}\label{diff1}
     \Delta S_A^{(n)}&=\frac{1}{1-n}\log\frac{\langle\mathcal{O}(\textbf{x}_1,t_1)\mathcal{O}^{\dagger}(\textbf{x}_2,t_2)\cdots \mathcal{O}_{2n}(\textbf{x}_{2n},t_{2n})\rangle}{(\langle\mathcal{O}(\textbf{x}_1,t_1)\mathcal{O}^{\dagger}(\textbf{x}_2,t_2)\rangle)_{\Sigma_1}^n }
 \end{align}
 Here path integral over $n$-sheeted manifold $Z_{\Sigma_n}$ is expressed in terms of the $2n$-point function on the replica surface where each sheet carries two operators located at different points. We will evaluate explicitly $\Delta S^{(n)}_A$ for scalar and free Maxwell theories.
 \subsection{Conformal scalar in $d=2$ dimension}\label{3}
 In this section we revisit the analysis of pseudo-entropy of conformal scalar in $d=2$ dimension \cite{Nakata:2020luh}. We compute the difference between the pseudo \Re entropy with the \Re entropy of the ground state.
 
 We begin by considering a massless scalar field theory in Euclidean $2$-dimesion with the co-ordinate $w=x+i \tau$. We create two states by acting two operators on the vacuum at the same spatial points but in different Euclidean times $a$ and $a'$.
 \begin{align}
     \begin{split}
        | \psi_1\rangle&=e^{-a H_{\rm{CFT}}}\mathcal{O}(x)|0\rangle\\
        | \psi_2\rangle&=e^{-a' H_{\rm{CFT}}}\mathcal{O}(x)|0\rangle
     \end{split}
 \end{align}
 For simplicity, let us begin by computing the variation of second pseudo-entropy $\Delta S^{(2)}_A$ explicitly.
 In \cite{Nakata:2020luh}, $\Delta S_{A}^{(2)}$ is computed as a function of the center of the sub-systems, and the inserted operators were kept at fixed spatial points. Here we will investigate $\Delta S_{A}^{(2)}$ as a function of the spatial insertion of the operators. Note that, we have translational invariance along the spatial direction and therefore $\Delta S^{(2)}_A$ should remain the same if we vary the center of the subsystems or vary the spatial positions of the operators.
 
\begin{figure}[htb]
\centering
\begin{subfigure}{.5\textwidth}
  \centering
  \includegraphics[width=1\linewidth]{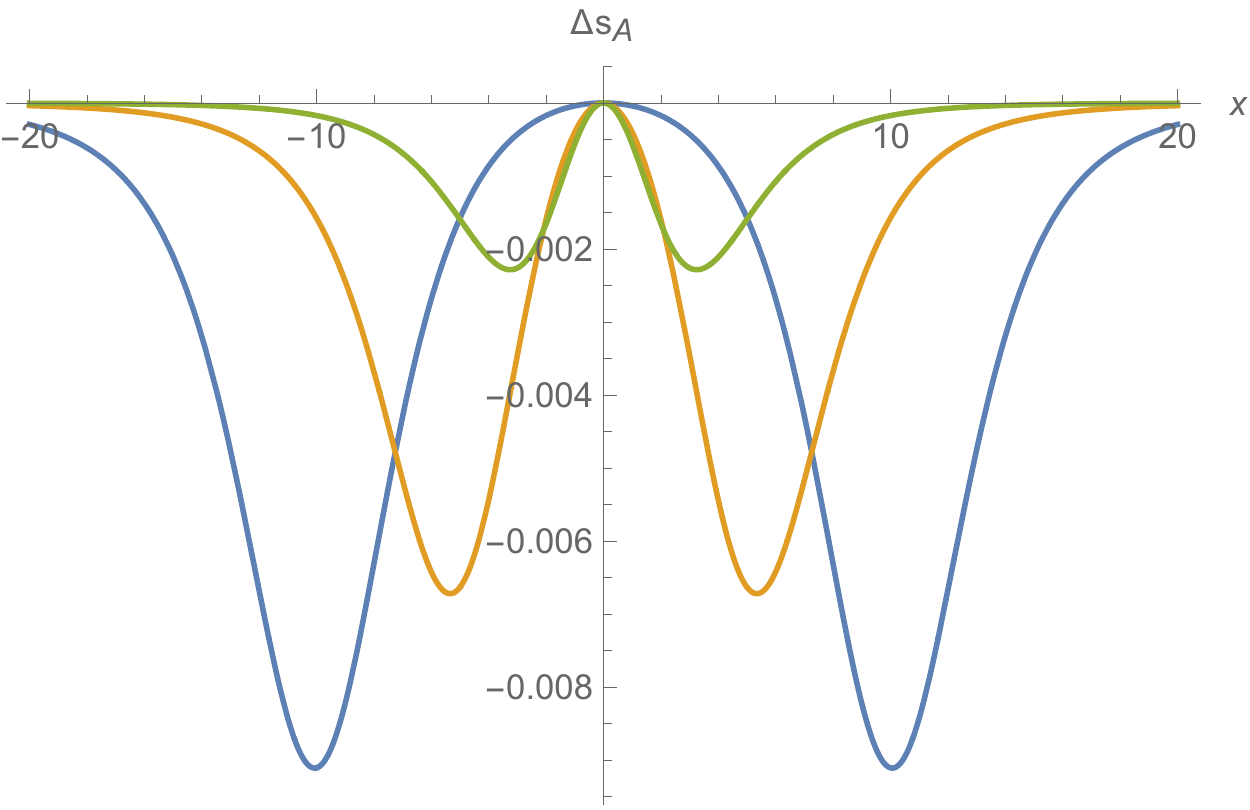}
  \caption{ $\Delta S^{(2)}_A$ obtained by varying the subsytem size.}
  \label{fig:sub12}
\end{subfigure}%
\begin{subfigure}{.5\textwidth}
  \centering
  \includegraphics[width=1\linewidth]{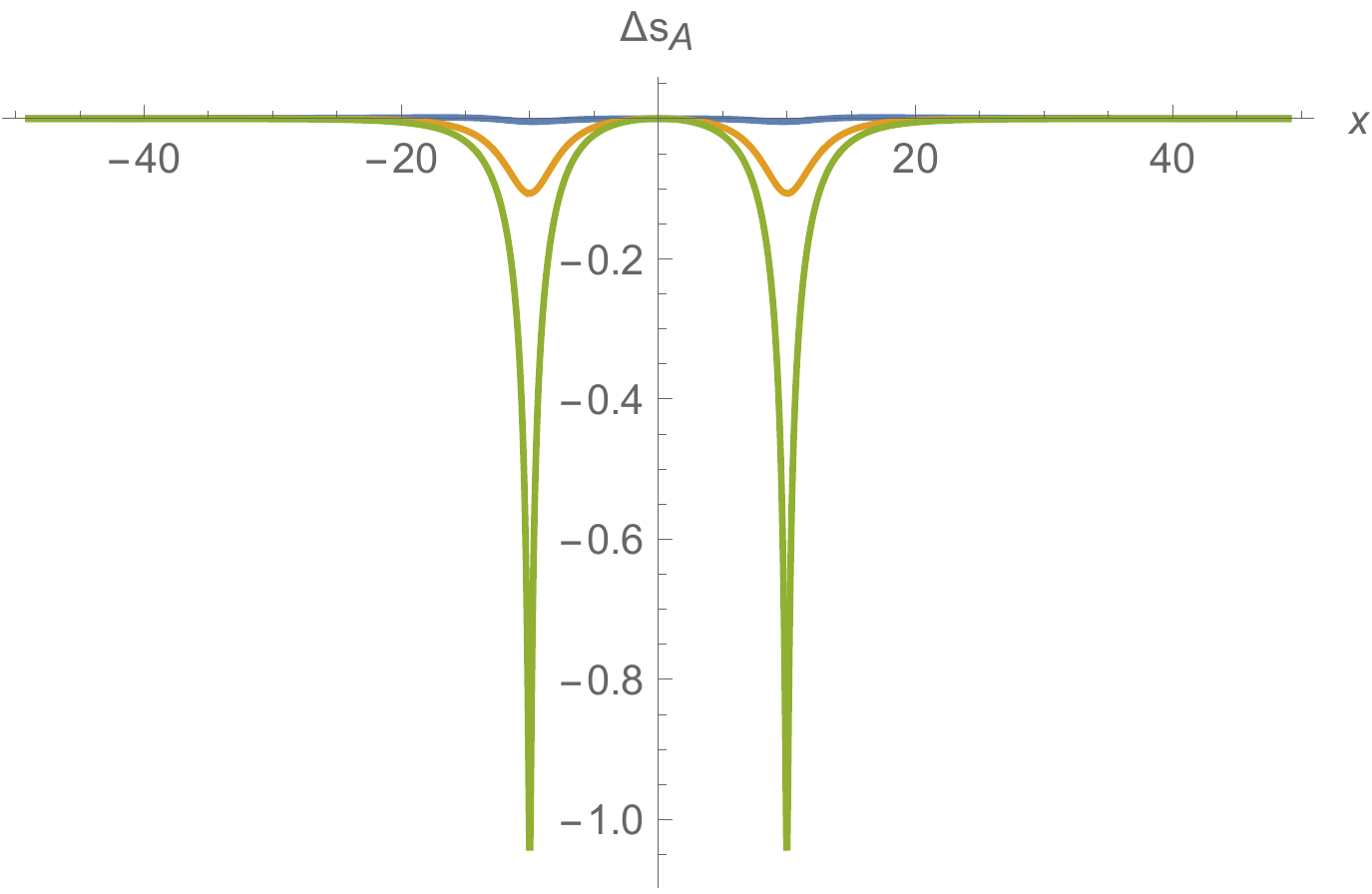}
  \caption{$\Delta S^{(2)}_A$ obtained by varying $a$ and $a'$.}
  \label{fig:sub22}
\end{subfigure}
\label{fig:test1}
\caption{The first plot shows the variation of $\Delta S^{(2)}_A$ with respect to subsytem size with fixed UV cutoffs.  Blue line: $\ell=20$, orange line: $\ell=10$, green line : $\ell=4$. The second plot shows the variation of  $\Delta S^{(2)}_A$  with respect to UV cutoffs at a fixed subsytem size of $\ell=20$. Blue
line: $a = 4$, $a' = 6$, orange line: $a = 2$, $a' = 8$, green line: $a = 0.1$, $a' = 9.9$. }
\end{figure}

 In the path integral picture, the operators are inserted at
 \begin{align}
     (w_1,\bar{w}_1)=(x-i a,x+i a),\quad\quad(w_2,\bar{w}_2)=(x-i a',x+ia')
 \end{align}
 We choose the  operator 
$
     \mathcal{O}=e^{\frac{i}{2}\phi}+e^{-\frac{i}{2}\phi}
 $, with conformal dimension $h=\bar{h}=\frac{1}{8}$.
 The variation of the second pseudo-\Re entropy becomes
 \begin{align}
     \Delta S_A^{(2)}&=-\log\frac{\langle\mathcal{O}(w_1,\bar{w}_1)\mathcal{O}^{\dagger}(w_2,\bar{w}_2)\mathcal{O}(w_3,\bar{w}_3)\mathcal{O}^{\dagger}(w_4,\bar{w}_4)\rangle}{\left(\langle\mathcal{O}(w_1,\bar{w}_1)\mathcal{O}^{\dagger}(w_2,\bar{w}_2)\rangle\right)^2}
 \end{align}
 So, the expression of $\Delta S^{(2)}_A$ involves the four and two-point functions of the operator $
     \mathcal{O}=e^{\frac{i}{2}\phi}+e^{-\frac{i}{2}\phi}
 $ on the replica surface. 
 To compute the four and two-point functions on the replica surface, one uses the conformal mapping
 \begin{align}\label{uni1}
     z&=\left(\frac{w-u}{w-v}\right)^{\frac{1}{2}}
 \end{align}
 This uniformization map takes branched cover $\Sigma_2$ to a plane. Note that, $u$ and $v$ are the end points of the subsystems which are held fixed. Since it is a free theory, one can evaluate the four-point functions easily using the Wick contraction of the operators. For $n=2$, two-point function  is given by \cite{Nozaki:2014uaa}
 \begin{align}
     \langle \phi(z_1,\bar{z}_1)\phi(z_2,\bar{z}_2)\rangle&=-\frac{1}{2}\log|z_1^{\frac{1}{2}}-z_2^{\frac{1}{2}}|+\frac{1}{2}\log\left(\frac{|z_1|^{-\frac{1}{2}}}{2}\right)+\frac{1}{2}\log\left(\frac{|z_2|^{-\frac{1}{2}}}{2}\right).
 \end{align}
 We evaluate the four-point functions and express $\Delta S^{(2)}_A$ as  a function of the cross-ratio
 \begin{align}
     \Delta S_A^{(2)}&=\log\frac{2}{1+|\eta|+|1-\eta|}.
 \end{align}
 The cross-ratio $\eta$ is given by
 \begin{align}
     \eta&=\frac{(z_1-z_2)(z_3-z_4)}{(z_1-z_3)(z_2-z_4)},
 \end{align}
 where $z_i$'s can be obtained from the relation \eqref{uni1} and $u,v$ are kept fixed. We evaluate $\Delta S_A^{(2)}$ as a function of the insertion of the operators. Figure \eqref{fig:sub12} shows $\Delta S_A^{(2)}$ for different sub-system size $u-v=\ell$, keeping $a$ and $a'$ fixed. We plot $\Delta S_A^{(2)}$ for $\ell=20$, $\ell=10$, $\ell=4$ and observe that $\Delta S_A^{(2)}$ picks up a sharp negative value when the insertion point becomes very close to the edges of the subsystems. Since we have translational invariance along the spatial line, one can also vary the center of the subsystems $x=\frac{u+v}{2}$ and observe the similar behavior of $\Delta S_A^{(2)}$ \cite{Nakata:2020luh}.
 
Therefore second pseudo \Re entropy is mostly zero, except at the points where operators become very close to the edges of the sub-systems.

This property can be understood in the language of entanglement swapping \cite{Nakata:2020luh}. When the spatial positions of the operators become close to the boundary, the system exhibits entanglement swapping. But entanglement swapping does not occur in the case where both the operators are located in one of the subsystems. In this case, there is no contribution to the $\Delta S^{(2)}_A$. Therefore, pseudo \Re entropy becomes the same as the ground state \Re entropy. It was also proved that the pseudo entropy is always greater than the original entanglement entropy of each state for 2-qubit systems but this is not true in general for all systems with larger degrees of freedom \cite{Nakata:2020luh}. For example, in the four-qubit systems, thermofield double states, and two-coupled harmonic oscillators the pseudo entropy becomes smaller than the original entanglement entropy of each state. Therefore the monotonicity of the pseudo entropy is not a general feature for all systems rather one has to investigate case by case. In this paper, we mainly focus on the variation of pseudo entropy with respect to the ground state \Re entropy.
 \subsection{Conformal scalar in $d=4$ dimension}\label{4}
  In this section, we study the pseudo-entropy in conformal field theory in $d=4$ dimension. In particular, we are interested to evaluate the difference between the pseudo R\'enyi entropy and the R\'enyi entropy of the ground states, and the expression is given in \eqref{diff1}. In $d=2$ dimension, we observe that the quantity $\Delta S^{(2)}_A$ decreases sharply near the boundary of the subsystems. Therefore we want to investigate the property of $\Delta S^{(2)}_A$ and particularly how it behaves near the boundary of the subsystem in $d=4$ dimension. 
 \subsection*{Same operator different insertion}
  At a constant time slice we subdivide the space and restrict one of the subspaces in the region of $x>0$ which means the two subsystems are separated by $y-z$ plane.
  
 We now consider two excited states which are prepared by acting two same operators but at different points on the vacuum in $d=4$ dimension. 
 \begin{align}
 \begin{split}
     |\psi\rangle&=e^{-a H}\phi(x_1,y_1,z_1)|0\rangle\\
     |\chi\rangle&=e^{-a' H}\phi(x_2,y_1,z_1)|0\rangle
 \end{split}
 \end{align}
 Here $\phi(x,y,z)$ is the conformal primary operator with unit dimension and $a$ and $a'$ are the cutoffs to avoid the UV divergences which can also be thought of as Euclidean times. Since we want to study $\Delta S^{(2)}_A$ as a function of the distance from the boundary, we substitute $y_1=y_2$ and $z_1=z_2$, which means two operators are placed at the same points along the boundary of the subsystems. Therefore $\Delta S^{(2)}_A$ is a function of the Euclidean times and the transverse distance from the boundary located at $x=0$.
 For computational simplification, we use polar coordinates for the $t-x$ plane and the other $y-z$ plane remains in cartesian. Therefore we work with the following metric
 \begin{align}
     ds^2=dr^2+r^2d\theta^2+dy^2+dz^2
 \end{align}
 We now compute $\Delta S_{A}^{(2)}$,
 \begin{align}
     \Delta S^{(2)}_A&=-\log \frac{\langle \phi(r_1,\theta_1^{(1)},y_1,z_1)\phi(r_2,\theta_2^{(1)},y_1,z_1)\phi(r_1,\theta_1^{(2)},y_1,z_1)\phi(r_2,\theta_2^{(2)},y_1,z_1)\rangle}{\langle \phi(r_1,\theta_1^,y_1,z_1)\phi(r_2,\theta_2,y_1,z_1)\rangle^2_{\Sigma_1}}
 \end{align}
We analyze the pseudo-entropy with the same operators but inserted at two different points. Since $y$ and $z$ coordinates are the same, $r_1$ and $r_2$ can be written as
 \begin{align*}
     r_1=\sqrt{a^2+x_1^2}, \quad    r_2=\sqrt{a'^2+x_2^2}
 \end{align*}
 Also, the angle between two points are given by
 \begin{align}
     \cos(\theta_1-\theta_2)&=\frac{ a a'+x_1x_2}{r_1 r_2}
 \end{align}
 To evaluate the $\Delta S_A^{(2)}$, we compute the four-point function by using Wick-contraction
 \begin{align}
     &\langle \phi(r_1,\theta_1^{(1)},y,z)\phi(r_2,\theta_2^{(1)},y,z)\phi(r_1,\theta_1^{(2)},y,z)\phi(r_2,\theta_2^{(2)},y,z)\rangle=\nonumber\\
     &\quad\quad\quad\quad\quad\quad\langle\phi(r_1,\theta_1^{(1)},y,z)\phi(r_2,\theta_2^{(1)},y,z)\rangle\langle \phi(r_1,\theta_1^{(2)},y,z)\phi(r_2,\theta_2^{(2)},y,z)\rangle\nonumber\\
     &\quad\quad\quad\quad\quad\quad+\langle\phi(r_1,\theta_1^{(1)},y,z)\phi(r_2,\theta_2^{(2)},y,z)\rangle\langle \phi(r_2,\theta_2^{(1)},y,z)\phi(r_1,\theta_1^{(2)},y,z)\rangle\nonumber\\
     &\quad\quad\quad\quad\quad\quad+\langle\phi(r_1,\theta_1^{(1)},y,z)\phi(r_1,\theta_1^{(2)},y,z)\rangle\langle \phi(r_2,\theta_2^{(1)},y,z)\phi(r_2,\theta_2^{(2)},y,z)\rangle
 \end{align}
 The two-point functions of conformal primaries  on the replica surface are known \cite{Nozaki:2014hna}. The two-point function which involves the two operators on the same sheet is given by
 \begin{align}\label{ss}
      \langle\phi(r_1,\theta_1^{(1)},y,z)\phi(r_2,\theta_2^{(1)},y,z)\rangle&=\langle \phi(r_1,\theta_1^{(2)},y,z)\phi(r_2,\theta_2^{(2)},y,z)\rangle\nonumber\\ &=\frac{1}{(8\pi^{2})(r_{1}+r_{2})\left((r_{1}+r_{2})-2\sqrt{r_{1}r_{2}}\cos{\frac{(\theta_{1}-\theta_{2})}{2}}\right)}
      \end{align}
      Similarly we have the two-point function which involves the operators across the sheet. This two-point function can be obtained by shifting $\theta_2\rightarrow \theta_2+2\pi$ in the expression of the two-point function on the same sheet. Therefore we obtain the two-point functions across the sheets
      \begin{align}
       \langle\phi(r_1,\theta_1^{(1)},y,z)\phi(r_2,\theta_2^{(2)},y,z)\rangle&=\langle \phi(r_2,\theta_2^{(1)},y,z)\phi(r_1,\theta_1^{(1)},y,z)\rangle\nonumber\\ &=\frac{1}{(8\pi^{2})(r_{1}+r_{2})\left((r_{1}+r_{2})+2\sqrt{r_{1}r_{2}}\cos{\frac{(\theta_{1}-\theta_{2})}{2}}\right)}
       \end{align}
       We also require the two-point functions across the sheets but involving the same points. This can be obtained by taking the limit $r_1\rightarrow r_2$ and $\theta_1\rightarrow \theta_2+2\pi$ in the expression of the correlator on the same sheet \eqref{ss}. So the two-point functions involving the same points across the sheets are given by
       \begin{align}
       \begin{split}
           \langle\phi(r_1,\theta_1^{(1)},y,z)\phi(r_1,\theta_1^{(2)},y,z)\rangle&=\frac{1}{(64\pi^{2}r_1^2)}\\
           \langle \phi(r_2,\theta_2^{(1)},y,z)\phi(r_2,\theta_2^{(2)},y,z)\rangle&=\frac{1}{(64\pi^{2}r_2^2)}
       \end{split}
         \end{align}
        
 \subsection*{Calculation of pseudo-entropy}
 Given all the two-point functions one can compute the four-point function explicitly. We therefore write the four-point functions explicitly in terms of $r$ and $\theta$ variables
 \begin{align}\label{4ptsc}
& \langle \phi(r_1,\theta_1^{(1)},y,z)\phi(r_2,\theta_2^{(1)},y,z)\phi(r_1,\theta_1^{(2)},y,z)\phi(r_2,\theta_2^{(2)},y,z)\rangle=\nonumber\\
 &\qquad\qquad\qquad\qquad\qquad\frac{\frac{64 \left(r_2 r_1 \left(\cos \left(\theta _1-\theta _2\right)+5\right)+2 r_1^2+2 r_2^2\right)}{\left(r_1+r_2\right){}^2 \left(-\frac{1}{2} r_2 r_1 \left(\cos \left(\theta _1-\theta _2\right)-3\right)+r_1^2+r_2^2\right){}^2}+\frac{1}{r_1^2 r_2^2}}{4096 \pi ^4}
 \end{align}
 The two-point function on the $n=1$ sheet is also given by
 \begin{align}\label{2pts1}
     \langle \phi(r_1,\theta_1,y,z) \phi(r_2,\theta_2,y,z)\rangle_{\Sigma_1}&=\frac{1}{4\pi^2\left(r_1^2+r_2^2-2 r_1 r_2 \cos(\theta_1-\theta_2)\right)}
 \end{align}
 This is just the usual two-point function on the flat space which depends on the distance between two pints.
 Therefore the pseudo-entropy for $n=2$ can be obtained
 \begin{align}\label{dels2sc}
      \Delta S^{(2)}_A&=-\log\frac{\mathcal{N}}{\mathcal{D}},\quad\quad\mathcal{N}=\frac{\frac{64 \left(r_2 r_1 \left(\cos \left(\theta _1-\theta _2\right)+5\right)+2 r_1^2+2 r_2^2\right)}{\left(r_1+r_2\right){}^2 \left(-\frac{1}{2} r_2 r_1 \left(\cos \left(\theta _1-\theta _2\right)-3\right)+r_1^2+r_2^2\right){}^2}+\frac{1}{r_1^2 r_2^2}}{4096 \pi ^4}\nonumber\\
      \quad\quad\quad\quad\quad\quad\quad\quad\mathcal{D}&=\left(\frac{1}{4\pi^2\left(r_1^2+r_2^2-2 r_1 r_2 \cos(\theta_1-\theta_2)\right)}\right)^2
 \end{align}
 Here $\mathcal{N}$ is the four-point function given in \eqref{4ptsc} and $\mathcal{D}$ is the square of the two-point function on $n=1$ sheet which is given in \eqref{2pts1}.

We now substitute $x_1=x_2=x$ 
 and plot the variation of pseudo R\'enyi entropy as function of  center of the two operators $x=\frac{x_1+x_2}{2}$.
\begin{figure}[htb]
\centering
\begin{subfigure}{.5\textwidth}
  \centering
  \includegraphics[width=1\linewidth]{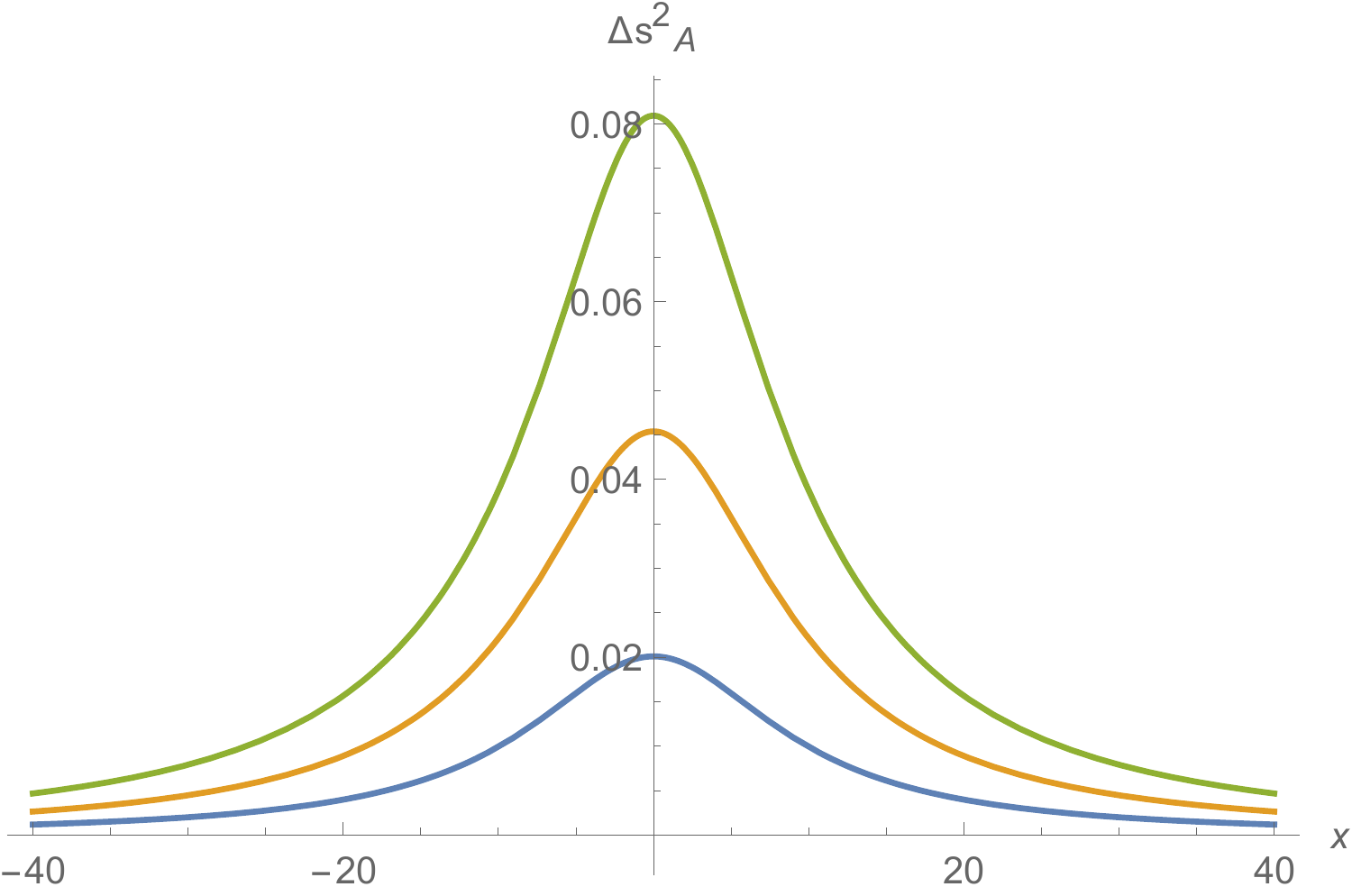}
  \caption{$\Delta S^{(2)}_A$ for $a\sim a'$. }
  \label{sc41}
\end{subfigure}%
\begin{subfigure}{.5\textwidth}
  \centering
  \includegraphics[width=1\linewidth]{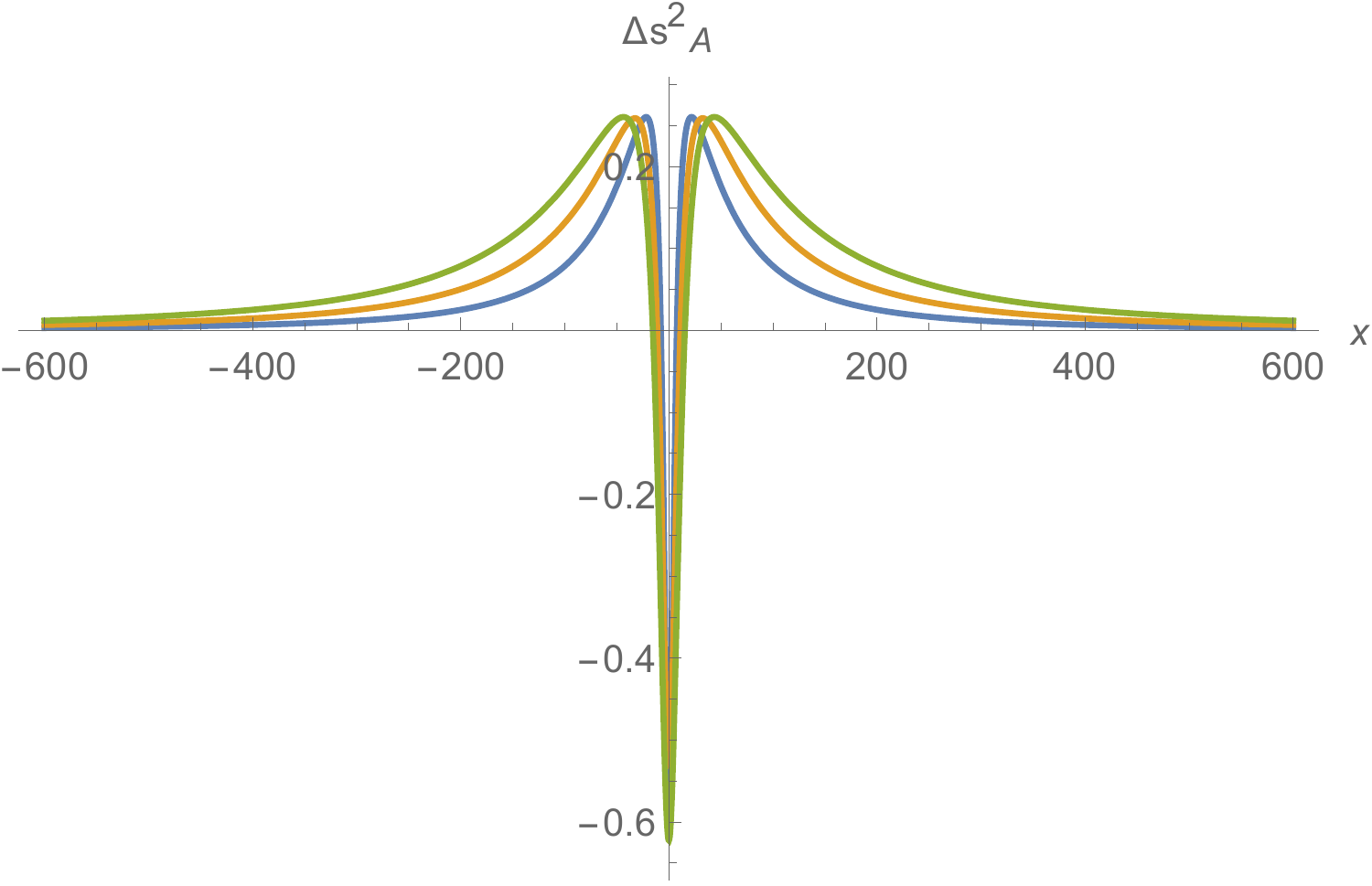}
  \caption{$\Delta S^{(2)}_A$ for $a\gg a'$.}
  \label{sc42}
\end{subfigure}
\label{scl42}
\caption{$\Delta S^{(2)}_A$ as a function of the center of the operators.}
\end{figure}
From the plot, we observe that $\Delta S^{(2)}_A$ increases as the center of the operators approaches  the boundary of the subsystems. This indicates  the entanglement swapping near the boundary of the subsystem.

Let us understand $\Delta S^{(2)}_A$ as a function of $x$. When the center of the two operators becomes very close to the boundary 
\begin{align}
    \lim_{x\rightarrow 0}\Delta S^{(2)}_A&=\log \left(\frac{256}{\frac{\left(a-a'\right)^4}{a^2 \left(a'\right)^2}+\frac{128 \left(a^2+6 a a'+\left(a'\right)^2\right)}{\left(a'+a\right)^2}}\right)+\mathcal{O}(x^2)+\cdots
\end{align}
We keep two different UV  cutoffs. Therefore it is expected that $\Delta S^{(2)}_A$ will take a finite positive value when $a$ and $a'$ are comparable which is $a\sim a'$. We define the ratio $\frac{a'}{a}=p$ because as we will see the near boundary behavior of $\Delta S^{(2)}_A$ will depend on this ratio.

When $p\sim 1$, which means the two UV cutoffs are comparable the leading behavior of the $  \lim_{x\rightarrow 0}\Delta S^{(2)}_A\sim \frac{1}{8} (p-1)^2>0$. In the first plot of, we keep $a\sim a'$ and observe that $\Delta S^{(2)}_A$ becomes a finite positive quantity near the boundary. In the plot \eqref{sc41} green line : $a=8$, $a'=12$; orange line $a=7$ , $a'=13$ and blue line $a=6$ and $a'=14$.

When $p\sim 0$, which means one of the UV cutoffs are negligible in compared to the other, the leading behavior of $  \lim_{x\rightarrow 0}\Delta S^{(2)}_A\sim(\log p^2)<0$. In the second plot \eqref{sc42} we consider the case where $a\gg a'$. We observe that $\Delta S^{(2)}_A$ becomes sharply negative near the boundary. In the plot, green line : $a=100$, $a'=5$; orange line $a=150$ , $a'=8$ and blue line $a=200$ and $a'=10$. 

But in both cases, $\Delta S^{(2)}_A$ changes significantly near the boundary of the subsystems due the transition in the entanglement. In the large $x$ limit where the center of the operators is far away from the boundary, $\Delta S^{(2)}_A$ becomes negligible.
\begin{align}
    \lim_{x\rightarrow \infty}\Delta S^{(2)}_A&=\frac{(a-a')^2}{8x^2}+\mathcal{O}(\frac{1}{x^4})+\cdots.
\end{align}
This feature is similar to the case of conformal scalar in $d=2$ dimension. When the operators are far away from the boundary there is no contribution to the $\Delta S^{(2)}_A$ indicating the fact that pseudo \Re entropy becomes the same as the ground state \Re entropy.
\subsection*{Real-time behavior}
Real-time behavior of $\Delta S^{(2)}_A$ can be evaluated by substituting $a=-i t-\epsilon$ and $a'=-i t+\epsilon$ in the expression given in \eqref{dels2sc}. Here $t$ is the real-time and $\epsilon$ is a small positive real number used to avoid the divergence. We observe that $\Delta S^{(2)}_A$ becomes $\log 2$ in the large time. This has been noted earlier in the context of local quench by a scalar primary operator in $d=4$ dimension \cite{Nozaki:2014hna,Nozaki:2014uaa,Nozaki:2015mca,Nozaki:2016mcy}. Therefore our computation of $\Delta S^{(2)}_A$ in \eqref{dels2sc} provides a good consistency checks in the real-time framework where one interprets it as a transition in the entanglement due to local quench of a scalar primary operator.

 \begin{figure}[htp]
    \centering
    \includegraphics[width=8cm]{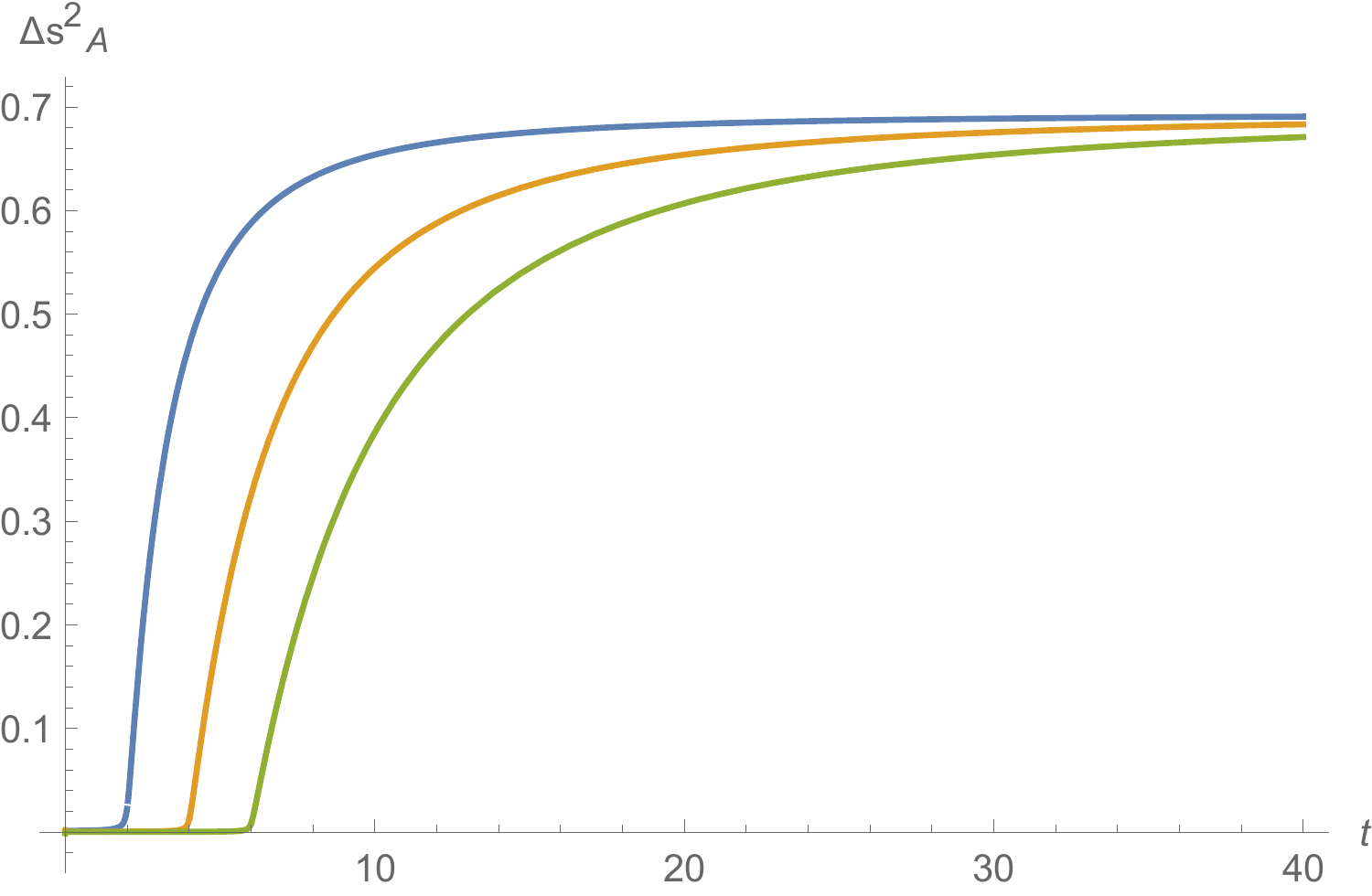}
    \caption{Real-time evolution of $\Delta S^{(2)}_A$. Blue line : $x=2$, ; orange line $x=4$;  green line $x=6$. We keep $\epsilon=0.1$ in all cases.}
    \label{pseudos}
\end{figure}
From the plot, we observe that $\Delta S^{(2)}_A=0$ when $t<x$ and it starts increasing immediately after $t=x$ and finally approaches to $\log 2$. This can be understood in the language of relativistic propagation of quasi-particles \cite{Nozaki:2014hna}. One can decompose  the scalar field $\phi=\phi_L+\phi_R$ where $\phi_L$ and $\phi_R$ corresponds to the left($x<0$) and right ($x>0$) moving modes. The entanglement between two modes kicks in at $t\geq x$ and they get maximally entangled at the large time.
We calculate the large time behavior,
\begin{align}
  \lim_{t\rightarrow \infty}  \Delta S^{(2)}_A&=\log 2-\frac{x^2}{t^2}+\mathcal{O}(\frac{1}{t^4}).
\end{align}
\section{Free Maxwell field in $d=4$ dimension}\label{5}
In this section, we evaluate the pseudo-entropy of the free Maxwell field in $d=4$ dimension. The field strength can be used to create excitations. We create excitations by the same components of the field strengths with different cutoffs. We also use the different components of the field strengths to create different states. We will follow the procedure developed in the previous section to compute $\Delta S^{(2)}_A$ analytically. In $d=4$ dimension, the free Maxwell theory is conformal, and therefore all the two-point functions and four-point functions can be computed exactly. But we are using the replica trick and therefore all the two-point functions have to be computed on the replica surface which was introduced in \cite{David:2020mls}.

We know that the $U(1)$ theory is gauge invariant under the transformation
      \begin{equation}
      A_\mu \rightarrow A_\mu + \partial_\mu \epsilon,
      \end{equation}
  where $\epsilon$ is the gauge parameter.    We can use the covariant gauge condition to fix the gauge
      \begin{equation}\label{covcond}
      \partial^\mu A_\mu =0.
      \end{equation}
       The equations of motion in the covariant gauge becomes
      \begin{equation}\label{maxeom}
      \nabla^2 A_\mu =0.
      \end{equation}
      Therefore under the gauge transformation
      \begin{equation}
      A_\mu' = A_\mu + \partial_\mu \epsilon, \qquad  {\rm with }\quad \Box \epsilon =0.
      \end{equation}
      Given a gauge potential which satisfies (\ref{covcond}) and (\ref{maxeom}) one can make 
      a further gauge transformation  so that 
      \begin{equation}\label{maxgcond}
      \partial^a A_a' =0, \qquad \partial^i A_{i}' =0  \quad a \in \{ r, \theta\} , i \in \{y, z\}.
      \end{equation}
      These  two gauge restrictions acting on the gauge potential can be done
         by choosing the gauge  transformation  to be
      \begin{equation}
      \epsilon  = - \frac{\partial^i A_i }{\nabla^2} , \qquad \nabla^2 = \partial_y^2 + \partial_z^2.
      \end{equation}
      Note that the gauge transformation also satisfies $\Box \epsilon=0$. Therefore, it is a valid choice of gauge. Note that \eqref{maxgcond}, are two gauge restrictions acting on two subspaces separately and gauge potential becomes transverse in both the subspaces.
      
      We need to evaluate the two and four-point functions on the replica surface.
      For this it is convenient to choose polar coordinates
      \begin{equation}
      ds^2 = dr^2  + r^2 d\theta^2 + ( dy)^2 + ( dz)^2.
      \end{equation}
      Here $\theta \sim \theta + 2\pi n $. This $n$ corresponds to the \Re parameter and one gets a periodicity in the $\theta$ coordinate after $2\pi n$ rotation.
      The two point function fo the gauge field  on the cone satisfying the gauge condition 
      (\ref{maxgcond}), is given by  \cite{Candelas:1977zza,David:2020mls}
      \begin{eqnarray}\label{corm1}
      G_{\mu\nu'} ( x,x')&=& \langle A_\mu ( x) A_{\nu'} ( x') \rangle.   \\ \nonumber
       G_{ab'} (x,x') &=&   \frac{P_a P_{b'} }{ \widehat\nabla^2}  G (x, x'),  \qquad
      G_{i j'}(x, x')    = \left[ \delta_{ij}  - \frac{\partial_i\partial_j}{ \nabla^2} \right] \tilde G(x, x') ,
      \\ \nonumber
     G_{a i'} (x,x') &=&  G_{i a'} ( x, x') =0.
      \end{eqnarray}
     All  two-point functions are transverse and $ G(x, x') $ is the scalar propagator on the cone which is given by 
      \begin{eqnarray}\label{sclg}
      G(x, x') &=& \frac{1}{ 4n\pi^2 r r' ( a - a^{-1} )} \frac{ a^{\frac{1}{n} }- a^{- \frac{1}{n}} }{
      a^{\frac{1}{n} }+  a^{- \frac{1}{n}} - 2 \cos \left( \frac{\theta - \theta'}{n} \right) }, \\ \nonumber
      \frac{a}{ 1+ a^2} &=& \frac{rr'}{ ( x^ i - x^{\prime i } )^2 + r^2 + r^{\prime 2} },\qquad r=\sqrt{t^2+x^2}, r'=\sqrt{t'^2+x'^2}.
      \end{eqnarray}
      and $P_a$ are defined as
\begin{eqnarray}\label{defpa}
P_a = \epsilon_{a b} g^{bc} \nabla_c , \qquad\qquad
\epsilon_{12} = -\epsilon_{21} = r,  \epsilon_{11} = \epsilon_{22} =0.
\end{eqnarray}
Note that, the scalar two-point function \eqref{sclg} on the replica surface is not invariant under translation in the $t$ and $x$ coordinate. So it is convenient to use the gauge we choose to write the correlators \eqref{corm1} on the replica surface. 
Using the gauge invariant two-point functions on the replica surface, we compute $\Delta S^{(2)}_A$ for different components of the field strength.
\subsection{Excitation by the same components of the field strength with different cutoffs}
The field strength is the gauge invariant operator and therefore
 different states prepared by the different components of the field strengths acting on vacuum remain gauge invariant. We choose two field strengths located at two different Euclidean times. We obtain pseudo \Re entropy for $n=2$ explicitly to study the properties of it.  
\subsection*{ Excitation by $F_{r\theta}$}We begin with the component $F_{r\theta}$. We prepare two states in the following way
\begin{align}
    \begin{split}
        |\psi_1\rangle&=e^{-\alpha H_{\rm{CFT}}}F_{r\theta}(x_1,y_1,z_1)|0\rangle\\
        |\psi_2\rangle &=e^{-\alpha' H_{\rm{CFT}}}F_{r\theta}(x_2,y_1,z_1)|0\rangle
    \end{split}
\end{align}
     We keep $y$ and $z$ coordinates of the operators the same and $\alpha$, $\alpha'$ are the two different cutoffs to avoid  UV divergence. The cutoffs $\alpha$ and $\alpha'$ distinguish two states. One can think of it as two operators located at two different Euclidean times. The subsystem is associated with $x>0$ region which means the $y-z$ plane separates the two subsystems. Therefore, two operators are separated only along the perpendicular direction from the boundary of the subsystems,
   We now evaluate $\Delta S_{A}^{(2)}$,
 \begin{align}
     \Delta S_{A}^{(2)}&=-\log \frac{\langle F_{r\theta}(r_1,\theta_1^{(1)},y_1,z_1)F_{r\theta }(r_2,\theta_2^{(1)},y_1,z_1)F_{r\theta}(r_1,\theta_1^{(2)},y_1,z_1)F_{r\theta }(r_2,\theta_2^{(2)},y_1,z_1)\rangle}{\langle F_{r\theta}(r_1,\theta_1^,y_1,z_1)F_{r\theta }(r_2,\theta_2,y_1,z_1)\rangle^2_{\Sigma_1}}
 \end{align}
 To evaluate $\Delta S_{A}^{(2)}$, we need to have the four-point and two-point functions of the gauge invariant operator $F_{r\theta}$.
  
Using the definitions of the gauge invariant two-point functions, we compute
 \begin{align}\label{2ptFrth}
         \langle F_{r\theta}(x_{i_1}) F_{r\theta}(x_{i_2})\rangle&=\partial_{r_1}\partial_{r_2}\langle A_{\theta}  A_{\theta}\rangle+\partial_{\theta_1}\partial_{\theta_2}\langle A_r A_r\rangle -\partial_{r_1}\partial_{\theta_2}\langle A_{\theta} A_{r}\rangle-\partial_{\theta_1}\partial_{r_2}\langle A_{r} A_{\theta}\rangle\nonumber\\
         &=\Big[\partial_{r_1}\partial_{r_2}(\frac{P_1 P_1'}{\nabla^2})+\partial_{\theta_1}\partial_{\theta_2}(\frac{P_0P_0'}{\nabla^2})-\partial_{r_1}\partial_{\theta_2}(\frac{P_1 P_0'}{\nabla^2})-\partial_{r_2}\partial_{\theta_1}(\frac{P_0 P_1'}{\nabla^2})\Big] G(x_{i_1};x_{i_2})\nonumber\\
         &=-(r_1r_2)\left(\partial_{r_1}^2+\frac{1}{r_1}\partial_{r_1}+\frac{1}{r_1^2}\partial_{\theta_1}^2\right) G(x_{i_1};x_{i_2})
       \end{align}
       Here $ G(x_{i_1};x_{i_2})$ is the massless scalar Green's function on the replica surface in $d=4$ dimension and the definition of the operator $P_a$ is given in \eqref{defpa}. $P_a$ denotes the operator located at the first coordinate and $P_a'$ denotes the operator located at the second coordinate.
       To derive the last line in \eqref{2ptFrth}, we use the on-shell condition. 
       $$\left(\partial_{r_1}^2+\frac{1}{r_1}\partial_{r_1}+\frac{1}{r_1^2}\partial_{\theta_1}^2+\nabla^2\right) G(x_{i_1};x_{i_2})=0,\qquad\qquad\qquad \nabla^2=\partial_{y_1}^2+\partial_{z_1}^2.$$
       It is now easy to compute the four-point function using the Wick contraction. The four-point function involves the correlators on the same sheets as well as the correlators across the  sheets. The two-point function across the sheet can be obtained by shifting $\theta_2\rightarrow \theta_2+2\pi$ in the expression of the correlators on the same sheet. This follows exactly the same pattern we observed in evaluating the four-point function of the scalar field in $d=4$ dimension. Therefore, the four-point function is given by
       \begin{align}\label{4ptFrth}
     &  \langle F_{r\theta}(r_1,\theta_1^{(1)},y_1,z_1)F_{r\theta }(r_2,\theta_2^{(1)},y_1,z_1)F_{r\theta}(r_1,\theta_1^{(2)},y_1,z_1)F_{r\theta }(r_2,\theta_2^{(2)},y_1,z_1)\rangle=\nonumber\\
     & \qquad\qquad\qquad\qquad \langle F_{r\theta}(r_1,\theta_1^{(1)},y_1,z_1)F_{r\theta }(r_2,\theta_2^{(1)},y_1,z_1)\rangle\langle F_{r\theta}(r_1,\theta_1^{(2)},y_1,z_1)F_{r\theta }(r_2,\theta_2^{(2)},y_1,z_1)\rangle+\nonumber\\
     &\qquad\qquad\qquad\qquad\langle  F_{r\theta}(r_1,\theta_1^{(1)},y_1,z_1)F_{r\theta}(r_1,\theta_1^{(2)},y_1,z_1)\rangle \langle F_{r\theta }(r_2,\theta_2^{(1)},y_1,z_1)F_{r\theta }(r_2,\theta_2^{(2)},y_1,z_1) \rangle +\nonumber\\
     &\qquad\qquad\qquad\qquad \langle F_{r\theta}(r_1,\theta_1^{(1)},y,z)F_{r\theta }(r_2,\theta_2^{(2)},y_1,z_1)\rangle\langle F_{r\theta }(r_2,\theta_2^{(1)},y_1,z_1)F_{r \theta}(r_1,\theta_1^{(2)},y_1,z_1)\rangle \nonumber\\
     &=\frac{r_1^2 r_2^2 \left(\sqrt{r_1 r_2} \left(-\cos \left(\frac{1}{2} \left(\theta _1-\theta _2\right)\right)\right)+r_1+r_2\right){}^2}{4 \pi ^4 \left(r_1+r_2\right){}^6 \left(-2 \sqrt{r_1 r_2} \cos \left(\frac{1}{2} \left(\theta _1-\theta _2\right)\right)+r_1+r_2\right){}^4}+\left(-\frac{3}{256 \pi ^2 r_1^2}\right)\left(-\frac{3}{256 \pi ^2 r_2^2}\right)\nonumber\\
     &\quad\qquad\qquad+\frac{r_1^2 r_2^2 \left(\sqrt{r_1 r_2} \cos \left(\frac{1}{2} \left(\theta _1-\theta _2\right)\right)+r_1+r_2\right){}^2}{4 \pi ^4 \left(r_1+r_2\right){}^6 \left(2 \sqrt{r_1 r_2} \cos \left(\frac{1}{2} \left(\theta _1-\theta _2\right)\right)+r_1+r_2\right){}^4}
    \end{align}
    We also  compute the two-point function on $n=1$ sheet
    \begin{align}\label{2ptFrth1}
        \langle F_{r\theta}(r_1,\theta_1^{(1)},y,z)F_{r\theta }(r_2,\theta_2^{(1)},y,z)\rangle_{\Sigma_1}&=-(r_1r_2)\left(\partial_{r_1}^2+\frac{1}{r_1}\partial_{r_1}+\frac{1}{r_1^2}\partial_{\theta_1}^2\right)G_{n=1}(x_{i_1};x_{i_2})\nonumber\\
        &=-\frac{r_1 r_2}{\pi ^2 \left(-2 r_2 r_1 \cos \left(\theta _1-\theta _2\right)+r_1^2+r_2^2\right){}^2}
    \end{align}
    Note that, $\Delta S^{(2)}_A$ becomes a function of $r$ and $\theta$ only because we keep the operators at the same $y$ and $z$ coordinates and separated them along the transverse direction from the boundary. We write $r_1$ and $r_2$
 \begin{align}\label{r1r2}
     r_1=\sqrt{\alpha^2+x_1^2}, \quad    r_2=\sqrt{\alpha'^2+x_2^2}
 \end{align}
 Also, the angle between two points are given by
 \begin{align}\label{cth}
     \cos(\theta_1-\theta_2)&=\frac{ \alpha \alpha'+x_1x_2}{r_1 r_2}
 \end{align} 
With two and four-point functions on the replica surface, we can obtain $\Delta S^{(2)}_A$. It becomes a function of the ratio of the four-point function and the square of the two-point function on the replica surface.
\begin{align}
    \Delta S^{(2)}_A&=-\log \frac{\mathcal{N}_1}{\mathcal{D}_1},
\end{align}
where $\mathcal{N}_1 $ is given in \eqref{4ptFrth} and $\mathcal{D}_1$ is the square of the expression given in \eqref{2ptFrth1}. Now we substitute $x_1=x_2$ and plot $\Delta S^{(2)}_A$ as a function of the center of the operators $x=\frac{x_1+x_2}{2}$.
\begin{figure}[htb]
\centering
\begin{subfigure}{.5\textwidth}
  \centering
  \includegraphics[width=1\linewidth]{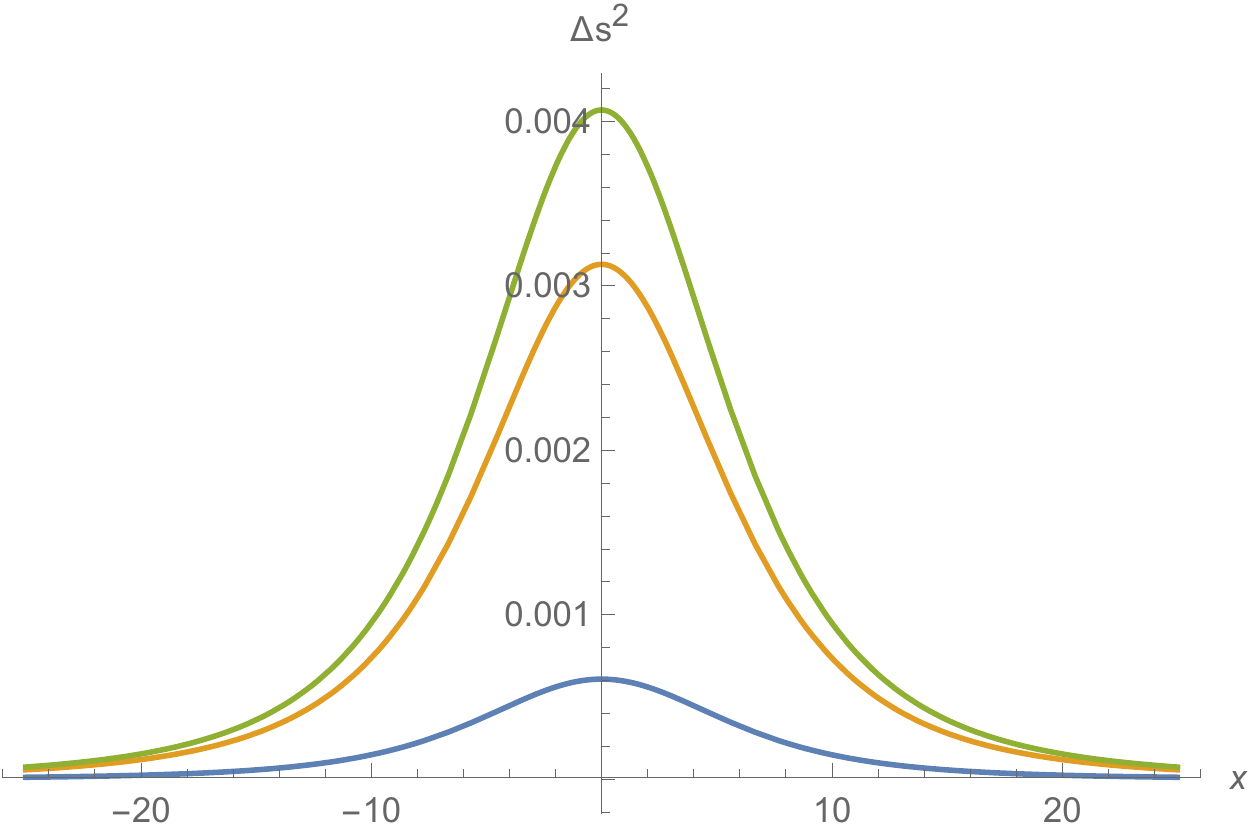}
  \caption{$\Delta S^{(2)}_A$ for $a\sim a'$. }
  \label{m1}
\end{subfigure}%
\begin{subfigure}{.5\textwidth}
  \centering
  \includegraphics[width=1\linewidth]{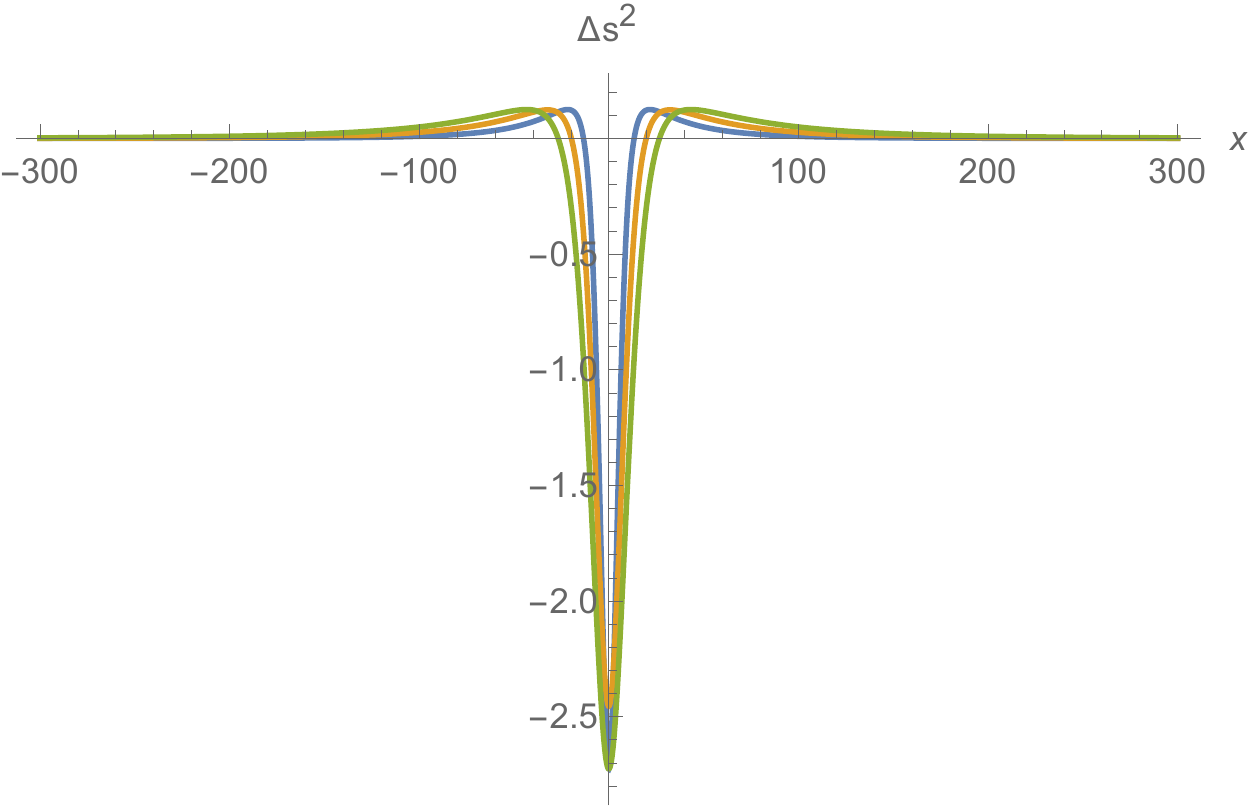}
  \caption{$\Delta S^{(2)}_A$ for $\alpha\gg \alpha'$.}
  \label{fig:sub22}
\end{subfigure}
\label{m2}
\caption{$\Delta S^{(2)}_A$ as a function of the center of the operators $F_{r\theta}$.}
\end{figure}

The two-point function of $F_{r\theta}$ on the replica surface is proportional to the scalar Laplacian (in $r$ and $\theta$ coordinates) acting on the Green's function. So one can expect a similar behavior of the pseudo entropy for the excited states prepared by field strength $F_{r\theta}$ acting on ground state at different Euclidean times. Let us investigate near boundary behavior of $\Delta S^{(2)}_A.$
\begin{align}
    \lim_{x\rightarrow 0}\Delta S^{(2)}_A&=\log \left(\frac{65536 \alpha ^4 \left(\alpha '\right)^4}{\left(\alpha -\alpha '\right)^8 \left(\frac{32768 \alpha ^4 \left(\alpha ^6+15 \alpha  \left(\alpha '\right)^5+\left(\alpha '\right)^6+15 \alpha ^5 \alpha '+27 \alpha ^4 \left(\alpha '\right)^2+42 \alpha ^3 \left(\alpha '\right)^3+27 \alpha ^2 \left(\alpha '\right)^4\right) \left(\alpha '\right)^4}{\left(\alpha -\alpha '\right)^8 \left(\alpha '+\alpha \right)^6}+9\right)}\right)+\cdots
\end{align}
It is clear that near boundary behavior of $\Delta S^{(2)}_A$ will depend on the ratio of the two Euclidean times  $p=\frac{\alpha'}{\alpha}$.

When $p\sim 1$, which means the two Euclidean times are comparable, the leading behavior of the $ \lim_{x\rightarrow 0}\Delta S^{(2)}_A\sim \frac{3}{128} (p-1)^4>0$. In the first plot of, we keep $\alpha\sim \alpha'$ and observe that $\Delta S^{(2)}_A$ becomes a finite positive quantity near the boundary. In the plot green line : $\alpha=8$, $\alpha'=12$; orange line $\alpha=7$ , $\alpha'=13$ and blue line $\alpha=6$ and $\alpha'=14$.

When $p \sim 0$, $\lim_{x\rightarrow 0} \Delta S^{(2)}_A \sim  \log (p)<0$. In the second plot we consider the case where $\alpha\gg \alpha'$. We observe that $\Delta S^{(2)}_A$ becomes sharply negative near the boundary. In the plot, green line : $\alpha=100$, $\alpha'=5$; orange line $\alpha=150$ , $\alpha'=8$ and blue line $\alpha=200$ and $\alpha'=10$. 

But in both cases, $\Delta S^{(2)}_A$ changes significantly near the boundary of the subsystems due to the transition in the entanglement. In the large $x$ limit where the center of the operators is far away from the boundary, $\Delta S^{(2)}_A$ becomes negligible. This is what we also observed in the case of the scalar field in $d=4$ dimension. Therefore, pseudo \Re entropy only differs from the ground state \Re entropy near the boundary of the subsystems.
 \subsection*{Real-time evolution}
  We also analyze the real-time evolution of $\Delta S^{2}_A$ for the excited states created by   $F_{r\theta}$ located at two different Euclidean times. To obtain the real-time expression, we substitute $\alpha=-i t-\epsilon$ and $\alpha'=-it+\epsilon$ in the expression of $\Delta S^{(2)}_A$, where $\epsilon$ is a small positive real number.
  \begin{figure}[htp]
    \centering
    \includegraphics[width=8cm]{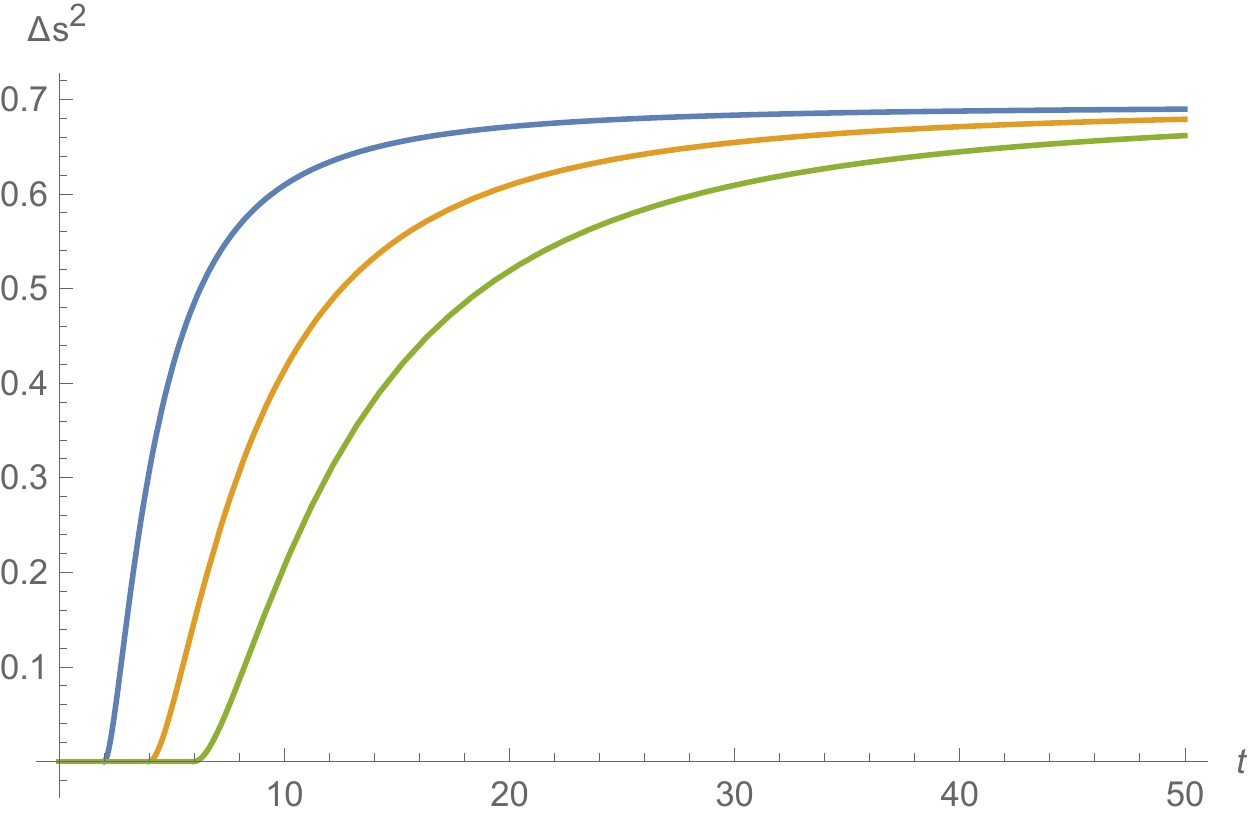}
    \caption{$\Delta S_A^{(2)}$ as a function for same components $F_{r\theta}$ of the field strength . Blue line : $x=2$, ; orange line $x=4$;  green line $x=6$. We keep $\epsilon=0.01$ in all cases.}
    \label{pseudos}
\end{figure}
We insert the operators at the same $y$ and $z$ co-ordinate. We fix the $x$ co-ordinate and observe the real-time dependence of $\Delta S^{(2)}_A$. 
We observe that in the large time $\Delta S_A^{(2)}$ reaches to $\log 2$ when the left and right moving states become maximally entangled \cite{Nozaki:2016mcy}.
\begin{align}
    \lim_{t\rightarrow \infty}\Delta S^{(2)}_A&=\log (2)-\frac{9 x^2}{4 t^2}+\cdots
\end{align}
Like the conformal scalar in $d=4$ dimension, $\Delta S^{(2)}_A$ remain zero till $t=x$. It starts growing after that and saturates to $\log 2$ at large time. But the growth of $\Delta S^{(2)}_A$ to reach the maximal entanglement differs from the scalar case. 
\subsection*{Excitation by $F_{yz}$}
 We  consider the case where the excitations are created by two same components of the field strength at different Euclidean times $\alpha$ and $\alpha'$. We choose the particular component to be $F_{yz}$.
 \begin{align}
    \begin{split}
        |\psi_1\rangle &=e^{-\alpha H_{\rm{CFT}}}F_{yz}(x_1,y_1,z_1)|0\rangle\\
        |\psi_2\rangle&=e^{-\alpha' H_{\rm{CFT}}}F_{yz}(x_2,y_1,z_1)|0\rangle
    \end{split}
    \end{align}
    We place the operators at the same $y$ and $z$ co-ordinates but in different $x$ coordinates. We want to study $\Delta S^{(2)}_A$  as a function of the center of the two operators. This is same as keeping the operators fixed and moving the center of the subsystem which is $x>0$ in this case.
 We now compute $\Delta S^{(2)}_{A}$,
 \begin{align}
     \Delta S^{(2)}_A&=-\log \frac{\langle F_{y z}(r_1,\theta_1^{(1)},y_1,z_1)F_{ yz}(r_2,\theta_2^{(1)},y,z)F_{yz}(r_1,\theta_1^{(2)},y_1,z_1)F_{ yz}(r_2,\theta_2^{(2)},y_1,z_1)\rangle}{\langle F_{yz}(r_1,\theta_1^,y,z)F_{yz}(r_2,\theta_2,y_1,z_1)\rangle^2_{\Sigma_1}}
 \end{align}
 To evaluate $\Delta S^{(2)}_A$, we require the two and four-point functions of $F_{yz}$ on the replica surface.  Using the definition of the two-point functions given in \eqref{corm1}, we obtain
 \begin{align}
     \langle F_{yz} F_{yz}\rangle&=\left(\partial_{y_1}\partial_{y_2}\langle A_z A_z\rangle+\partial_{z_1}\partial_{z_2}\langle A_y A_y\rangle\right)\nonumber\\
     &=-\frac{1}{2}(\partial_{y_1}^2+\partial_{z_1}^2)G(x_{i_1},x_{i_2})\nonumber\\
     &=\frac{1}{2}(\partial_{r_1}^2+\frac{1}{r_1}\partial_{\theta_1}+\frac{1}{r_1^2}\partial_{\theta_1}^2)G(x_{i_1},x_{i_2})
 \end{align}
In the second line, we use the translational invariance in the $y$ and $z$ coordinates and the isotropy in the $y-z$ plane. This can be checked very easily that
\begin{align}\label{isot}
    \partial_{y_1}^2G(x_{i_1},x_{i_2})=\partial_{z_1}^2 G(x_{i_1},x_{i_2})=\frac{1}{2}\nabla^2G(x_{i_1},x_{i_2}).
\end{align} 
In the last line we use the on-shell condition which is given by
 \begin{align}
     (\partial_{r_1}^2+\frac{1}{r_1}\partial_{\theta_1}+\frac{1}{r_1^2}\partial_{r_1}^2+\partial_{y_1}^2+\partial_{z_1}^2)G(x_{i_1},x_{i_2})=0
 \end{align} Here $G(x_{i_1},x_{i_2})$ is the scalar two-point function on the replica surface. 
  It is now easy to compute the four-point function using the two-point functions on the replica surface. The four-point function involves the correlator on the same sheet and correlators across the sheets as well.
  \begin{align}
     & \langle F_{y z}(r_1,\theta_1^{(1)},y_1,z_1)F_{ yz}(r_2,\theta_2^{(1)},y,z)F_{yz}(r_1,\theta_1^{(2)},y_1,z_1)F_{ yz}(r_2,\theta_2^{(2)},y_1,z_1)\rangle=\nonumber\\
      & \qquad\qquad\qquad\qquad \langle F_{yz}(r_1,\theta_1^{(1)},y_1,z_1)F_{yz }(r_2,\theta_2^{(1)},y_1,z_1)\rangle\langle F_{yz}(r_1,\theta_1^{(2)},y_1,z_1)F_{yz }(r_2,\theta_2^{(2)},y_1,z_1)\rangle+\nonumber\\
     &\qquad\qquad\qquad\qquad\langle  F_{yz}(r_1,\theta_1^{(1)},y_1,z_1)F_{yz}(r_1,\theta_1^{(2)},y_1,z_1)\rangle \langle F_{yz }(r_2,\theta_2^{(1)},y_1,z_1)F_{yz }(r_2,\theta_2^{(2)},y_1,z_1) \rangle +\nonumber\\
     &\qquad\qquad\qquad\qquad \langle F_{yz}(r_1,\theta_1^{(1)},y,z)F_{yz }(r_2,\theta_2^{(2)},y_1,z_1)\rangle\langle F_{yz }(r_2,\theta_2^{(1)},y_1,z_1)F_{yz}(r_1,\theta_1^{(2)},y_1,z_1)\rangle \nonumber\\
      &=\frac{1}{2}\Big[\frac{r_1 r_2 \left(\sqrt{r_1 r_2} \left(-\cos \left(\frac{1}{2} \left(\theta _1-\theta _2\right)\right)\right)+r_1+r_2\right){}^2}{4 \pi ^4 \left(r_1+r_2\right){}^6 \left(-2 \sqrt{r_1 r_2} \cos \left(\frac{1}{2} \left(\theta _1-\theta _2\right)\right)+r_1+r_2\right){}^4}+\left(-\frac{3}{256 \pi ^2 r_1^3}\right)\left(-\frac{3}{256 \pi ^2 r_2^3}\right)\nonumber\\
     &\quad\qquad\qquad+\frac{r_1 r_2 \left(\sqrt{r_1 r_2} \cos \left(\frac{1}{2} \left(\theta _1-\theta _2\right)\right)+r_1+r_2\right){}^2}{4 \pi ^4 \left(r_1+r_2\right){}^6 \left(2 \sqrt{r_1 r_2} \cos \left(\frac{1}{2} \left(\theta _1-\theta _2\right)\right)+r_1+r_2\right){}^4}\Big]
  \end{align}
    We also  compute the two-point function on $n=1$ sheet
    \begin{align}
        \langle F_{yz}(r_1,\theta_1^{(1)},y,z)F_{yz }(r_2,\theta_2^{(1)},y_1,z_1)\rangle_{\Sigma_1}&=\frac{1}{2}\left(\partial_{r_1}^2+\frac{1}{r_1}\partial_{r_1}+\frac{1}{r_1^2}\partial_{\theta_1}^2\right)G(x_{i_1};x_{i_2})_{n=1}\nonumber\\
        &=\frac{1}{2}\frac{1}{\pi ^2 \left(-2 r_2 r_1 \cos \left(\theta _1-\theta _2\right)+r_1^2+r_2^2\right){}^2}
    \end{align}
 $\Delta S^{(2)}_A$ is a function of $r$ and $\theta$ only because we keep the operators at the same $y$ and $z$ co-ordinates. Note that, two-point function of $F_{yz}$ is also proportional to the scalar Laplacian acting on the scalar Green's function in $d=4$ dimension. This reflects the duality between the field strengths $F_{r\theta}$ and $F_{yz}$ in Euclidean coordinates.
 \begin{align}
     F_{r\theta}(r)&=\frac{i}{2}\sqrt{g}F_{yz}(r).
 \end{align} Now we plot $\Delta S^{(2)}_A$ as a function of the center of the two operators.
\begin{figure}[htb]
\centering
\begin{subfigure}{.5\textwidth}
  \centering
  \includegraphics[width=1\linewidth]{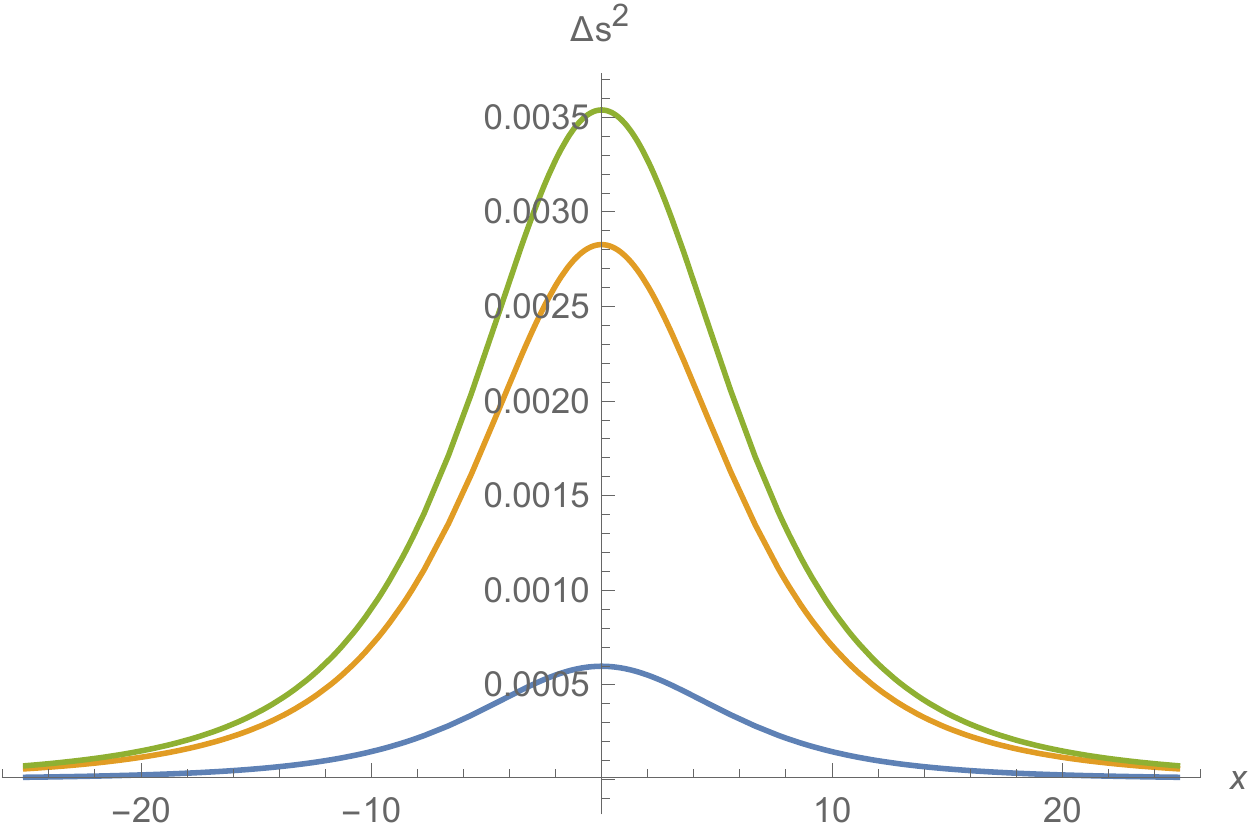}
  \caption{$\Delta S^{(2)}_A$ for $\alpha\sim \alpha'$. }
  \label{m3}
\end{subfigure}%
\begin{subfigure}{.5\textwidth}
  \centering
  \includegraphics[width=1\linewidth]{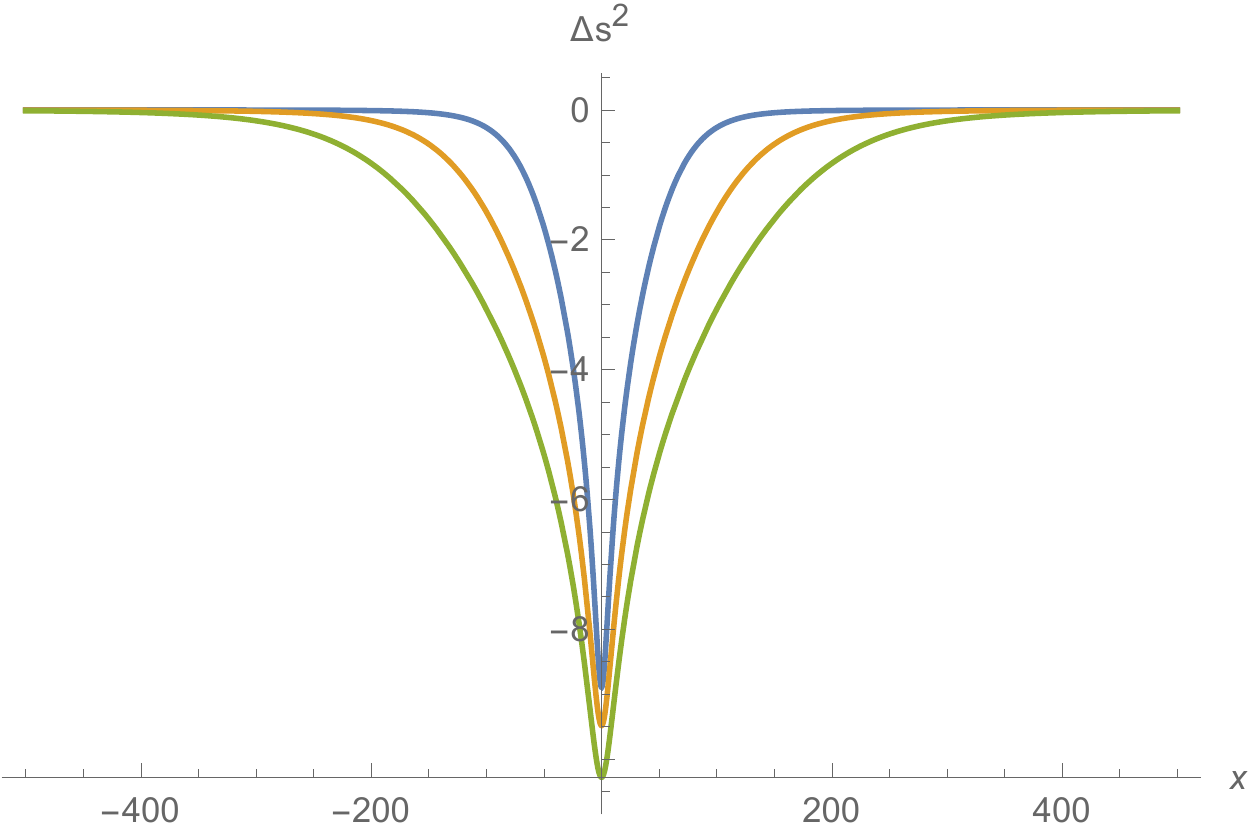}
  \caption{$\Delta S^{(2)}_A$ for $\alpha\gg \alpha'$.}
  \label{m4}
\end{subfigure}
\label{EE}
\caption{$\Delta S^{(2)}_A$ as a function of the center of the operators $F_{yz}$.}
\end{figure}

From the plot, we observe that $\Delta S^{(2)}_A$ changes significantly when the operators are very close to the boundary of the subsystems and there is no contribution to the $\Delta S^{(2)}_A$ far away from the boundary. This property is similar to the case of a scalar field in $d=4$ dimension.
We now investigate the near boundary behavior of $\Delta S^{(2)}_A$.
\begin{align}
    \lim_{x\rightarrow 0}\Delta S^{(2)}_A&=16 \log (2)-\log\Big(\left(\alpha -\alpha '\right)^8\big(\frac{16384 \left(-\sqrt{\alpha  \alpha '}+\alpha '+\alpha \right)^2}{\left(\alpha '+\alpha \right)^6 \left(-2 \sqrt{\alpha  \alpha '}+\alpha '+\alpha \right)^4}+\frac{9}{\alpha ^3 \left(\alpha '\right)^3}\nonumber\\
    &\qquad\qquad\qquad\qquad+\frac{16384 \left(\sqrt{\alpha  \alpha '}+\alpha '+\alpha \right)^2}{\left(\alpha '+\alpha \right)^6 \left(2 \sqrt{\alpha  \alpha '}+\alpha '+\alpha \right)^4}\big)\Big)
\end{align}
Evidently, the near boundary behavior of $\Delta S^{(2)}_A$ will depend on the ratio of two Euclidean times $p=\frac{\alpha'}{\alpha}$. When $\alpha\sim \alpha'$, $\Delta S^{(2)}_A \sim \frac{3}{128} (p-1)^4>0$. Therefore two comparable Euclidean times of the operators leads to a small finite positive $\Delta S^{(2)}_A$ near the boundary of subsystems. In the first plot \eqref{m3}, we keep $\alpha\sim \alpha'$ and observe that $\Delta S^{(2)}_A$ becomes a finite positive quantity near the boundary. In the plot green line : $\alpha=8$, $\alpha'=12$; orange line $\alpha=7$ , $\alpha'=13$ and blue line $\alpha=6$ and $\alpha'=14$.

In the $p \sim 0$, $\lim_{x\rightarrow 0} \Delta S^{(2)}_A \sim \log p^3<0$. In the second plot \eqref{m4}, we consider the case where $\alpha\gg \alpha'$. We observe that $\Delta S^{(2)}_A$ becomes sharply negative near the boundary. In the plot, green line : $\alpha=100$, $\alpha'=5$; orange line $\alpha=150$ , $\alpha'=8$ and blue line $\alpha=200$ and $\alpha'=10$. Similar to the scalar case, the variation of the pseudo entropy is significant near the boundary of the subsystems due to the entanglement swapping but far away from the boundary there is no transition in the entanglement and hence no contribution to $\Delta S^{(2)}_A.$
 \subsection*{Real-time evolution}
  We also analyze the real-time evolution of $\Delta S^{(2)}_A$ for the same components $F_{yz}$ of field strength. We substitute $\alpha=-i t-\epsilon$ and $\alpha'=-it+\epsilon$ in the expression of $\Delta S^{(2)}_A$, where $\epsilon$ is a small positive real number.
  \begin{figure}[htp]
    \centering
    \includegraphics[width=8cm]{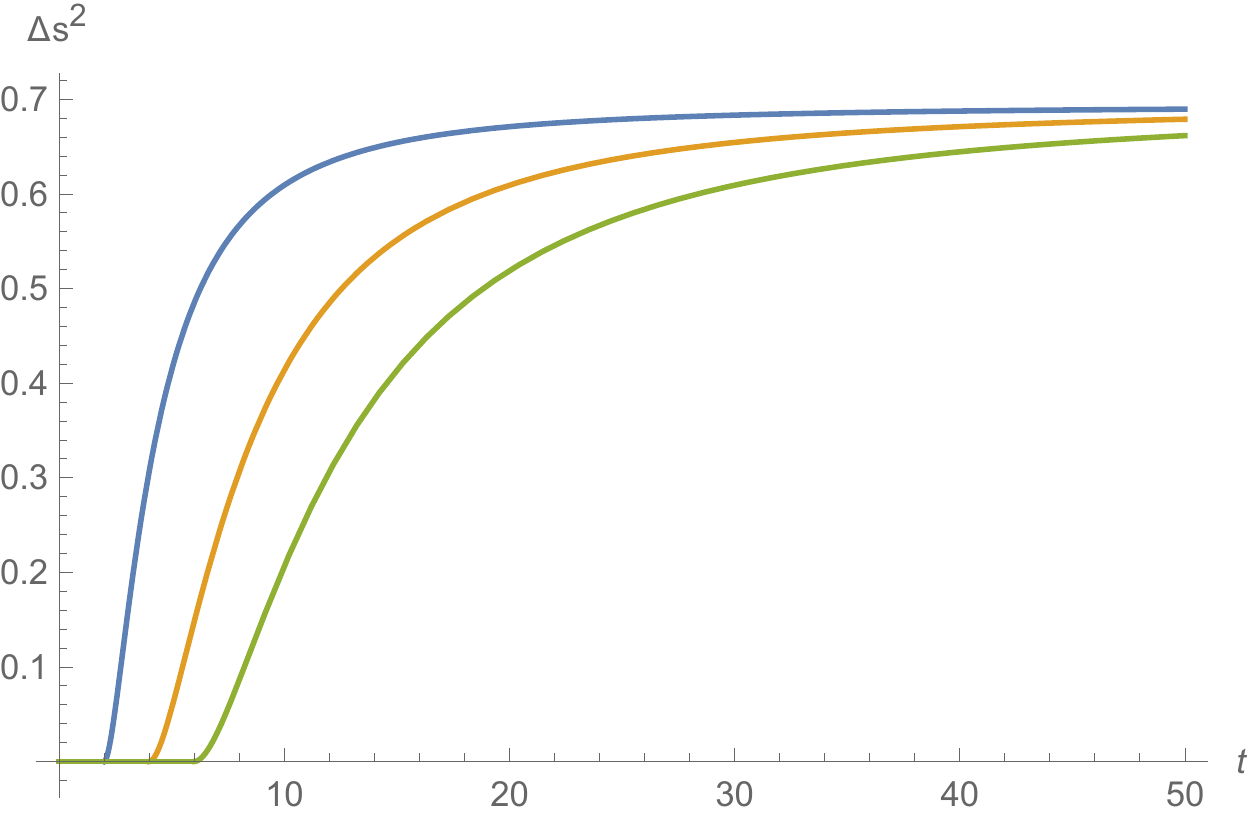}
    \caption{$\Delta S_A^{(2)}$ as a function for same components $F_{yz}$ of the field strength . Blue line : $x=2$, ; orange line $x=4$;  green line $x=6$. We keep $\epsilon=0.01$ in all cases.}
    \label{pseudos}
\end{figure}
We insert the operators at the same $y$ and $z$ co-ordinate. We fix the $x$ co-ordinate and observe the time depenence of $\Delta S^{(2)}_A$. 
We observe that in the large time $\Delta S_A^{(2)}$ reaches to $\log 2$.
\begin{align}
    \lim_{t\rightarrow \infty}\Delta S^{(2)}_A&=\log (2)-\frac{9 x^2}{4 t^2}+\cdots
\end{align}
Note that $\Delta S^{(2)}_A$ due to the local quench by the operator $F_{yz}$ is identical to that of $\Delta S^{(2)}_A$ by the operator $F_{r\theta}$ in the large $t$ limit. This is the consequence of the duality relation between the field strengths. This duality relation is reflected explicitly on the two-point functions and hence in $\Delta S^{(2)}_A$. The explicit time dependence of $\Delta S^{(2)}_A$ also remains same and saturates to $\log 2$ at $t\rightarrow \infty$ when two excited states created by $F_{yz}$ acting on the vacuum become maximally entangled.
\subsection*{Excitation created by $F_{ry}$}
We  consider the case where the excitations are created by two same components of the field strength at different Euclidean time $\alpha$ and $\alpha'$. In this case we choose the field strength to be $F_{ry}$.
 \begin{align}
    \begin{split}
        |\psi_1\rangle &=e^{-\alpha H_{\rm{CFT}}}F_{ry}(x_1,y_1,z_1)|0\rangle\\
        |\psi_2\rangle&=e^{-\alpha' H_{\rm{CFT}}}F_{ry}(x_2,y_1,z_1)|0\rangle
    \end{split}
    \end{align}
    We place the operators at the same $y$ and $z$ co-ordinates but in different $x$ coordinates. We want to study $\Delta S^{(2)}_A$  as a function of the center of the two operators. This is same as keeping the operators fixed and moving the center of the subsystem which is $x>0$ in this case. We now compute $\Delta S^{(2)}_{A}$,
 \begin{align}
     \Delta S^{2}_A&=-\log \frac{\langle F_{ ry}(r_1,\theta_1^{(1)},y_1,z_1)F_{ ry}(r_2,\theta_2^{(1)},y,z)F_{ry}(r_1,\theta_1^{(2)},y_1,z_1)F_{ ry}(r_2,\theta_2^{(2)},y_1,z_1)\rangle}{\langle F_{ry}(r_1,\theta_1^,y,z)F_{ry}(r_2,\theta_2,y_1,z_1)\rangle^2_{\Sigma_1}}
 \end{align}
 To evaluate $\Delta S^{2}_A$, we require the four and two-point functions of $F_{ry}$ on the replica surface.  Using the definition of the two-point functions given in \eqref{corm1}, we obtain
 \begin{align}\label{2ptFry}
     \langle F_{ry} F_{ry}\rangle&=\left(\partial_{r_1}\partial_{r_2}\langle A_y A_y\rangle+\partial_{y_1}\partial_{y_2}\langle A_r A_r\rangle \right)\nonumber\\
     &=\left(\partial_{r_1}\partial_{r_2}\frac{\partial_{z_1}^2}{\nabla^2}-\partial_{y_1}^2\frac{\partial_{\theta_1}\partial_{\theta_2}}{r_1r_2\nabla^2}\right)G(x_{i_1},x_{i_2})\nonumber\\
     &=\frac{1}{2}\left(\partial_{r_1}\partial_{r_2}-\frac{1}{r_1r_2}\partial_{\theta_1}\partial_{\theta_2}\right)G(x_{i_1},x_{i_2})
 \end{align}
 In the second line, we use the translational invariance of the scalar Green's function in the $y$ and $z$ coordinate and in the final line, we use the isotropic relation \eqref{isot} in the $y-z$ plane.
  Here $G(x_{i_1},x_{i_2})$ is the scalar two-point function on the replica surface. To evaluate $\Delta S^{(2)}$, we need the four-point function which we evaluate
  \begin{align}
        &  \langle F_{ry}(r_1,\theta_1^{(1)},y,z)F_{r y}(r_2,\theta_2^{(1)},y,z)F_{ry}(r_1,\theta_1^{(2)},y,z)F_{r y}(r_2,\theta_2^{(2)},y,z)\rangle=\nonumber\\
     & \qquad\qquad\qquad\qquad \langle F_{ry}(r_1,\theta_1^{(1)},y,z)F_{r y}(r_2,\theta_2^{(1)},y,z)\rangle\langle F_{ry}(r_1,\theta_1^{(2)},y,z)F_{r y}(r_2,\theta_2^{(2)},y,z)\rangle+\nonumber\\
     &\qquad\qquad\qquad\qquad \langle F_{ry}(r_1,\theta_1^{(1)},y,z)F_{r y}(r_2,\theta_2^{(2)},y,z)\rangle\langle F_{r y}(r_2,\theta_2^{(1)},y,z)F_{r y}(r_1,\theta_1^{(2)},y,z)\rangle \nonumber\\
     &\qquad\qquad\qquad\qquad\langle  F_{ry}(r_1,\theta_1^{(1)},y,z)F_{ry}(r_1,\theta_1^{(2)},y,z)\rangle \langle F_{r y}(r_2,\theta_2^{(1)},y,z)F_{r y}(r_2,\theta_2^{(2)},y,z) \rangle +\nonumber\\
     &=\Big(\frac{\left(r_1^2+6 r_2 r_1+r_2^2\right) \left(r_1 r_2 \cos \left(\theta _1-\theta _2\right)-3 \sqrt{r_1 r_2} \left(r_1+r_2\right) \cos \left(\frac{1}{2} \left(\theta _1-\theta _2\right)\right)\right)+r_1 r_2 \left(9 r_1^2+22 r_2 r_1+9 r_2^2\right)}{16 \pi ^2 r_1 r_2 \left(r_1+r_2\right){}^3 \left(-2 \sqrt{r_1 r_2} \cos \left(\frac{1}{2} \left(\theta _1-\theta _2\right)\right)+r_1+r_2\right){}^3}\Big)^2\nonumber\\
     &+\Big(\frac{\left(r_1^2+6 r_2 r_1+r_2^2\right) \left(r_1 r_2 \cos \left(\theta _1-\theta _2\right)+3 \sqrt{r_1 r_2} \left(r_1+r_2\right) \cos \left(\frac{1}{2} \left(\theta _1-\theta _2\right)\right)\right)+r_1 r_2 \left(9 r_1^2+22 r_2 r_1+9 r_2^2\right)}{16 \pi ^2 r_1 r_2 \left(r_1+r_2\right){}^3 \left(2 \sqrt{r_1 r_2} \cos \left(\frac{1}{2} \left(\theta _1-\theta _2\right)\right)+r_1+r_2\right){}^3}\Big)^2\nonumber\\
      &+ \left(\frac{3}{256\pi^2r_2^4}\right)\left(\frac{3}{256\pi^2r_1^4}\right)
  \end{align}
  The four-point function involves the correlator on the same sheet and the correlator across the sheets.
  We also evaluate the two-point function on $n=1$ sheet,
  \begin{align}\label{sh1Fry}
  \langle F_{ry}F_{ry}\rangle_{\Sigma_1}&=\frac{1}{2}\left(\partial_{r_1}\partial_{r_2}-\frac{1}{r_1r_2}\partial_{\theta_1}\partial_{\theta_2}\right)G(x_{i_1},x_{i_2})_{n=1}\nonumber\\
  &=\frac{2 r_1 r_2-\left(r_1^2+r_2^2\right) \cos \left(\theta _1-\theta _2\right)}{\pi ^2 \left(-2 r_2 r_1 \cos \left(\theta _1-\theta _2\right)+r_1^2+r_2^2\right){}^3}
  \end{align}$\Delta S^{(2)}_A$ is now a function of $r$ and $\theta$ only because we keep the operators at the same $y$ and $z$ co-ordinates. We substitute $x_1=x_2$ and  plot $\Delta S^{(2)}_A$ as a function of the center of the two operators $x=\frac{x_1+x_2}{2}.$
 \begin{figure}[htp]
    \centering
    \includegraphics[width=8cm]{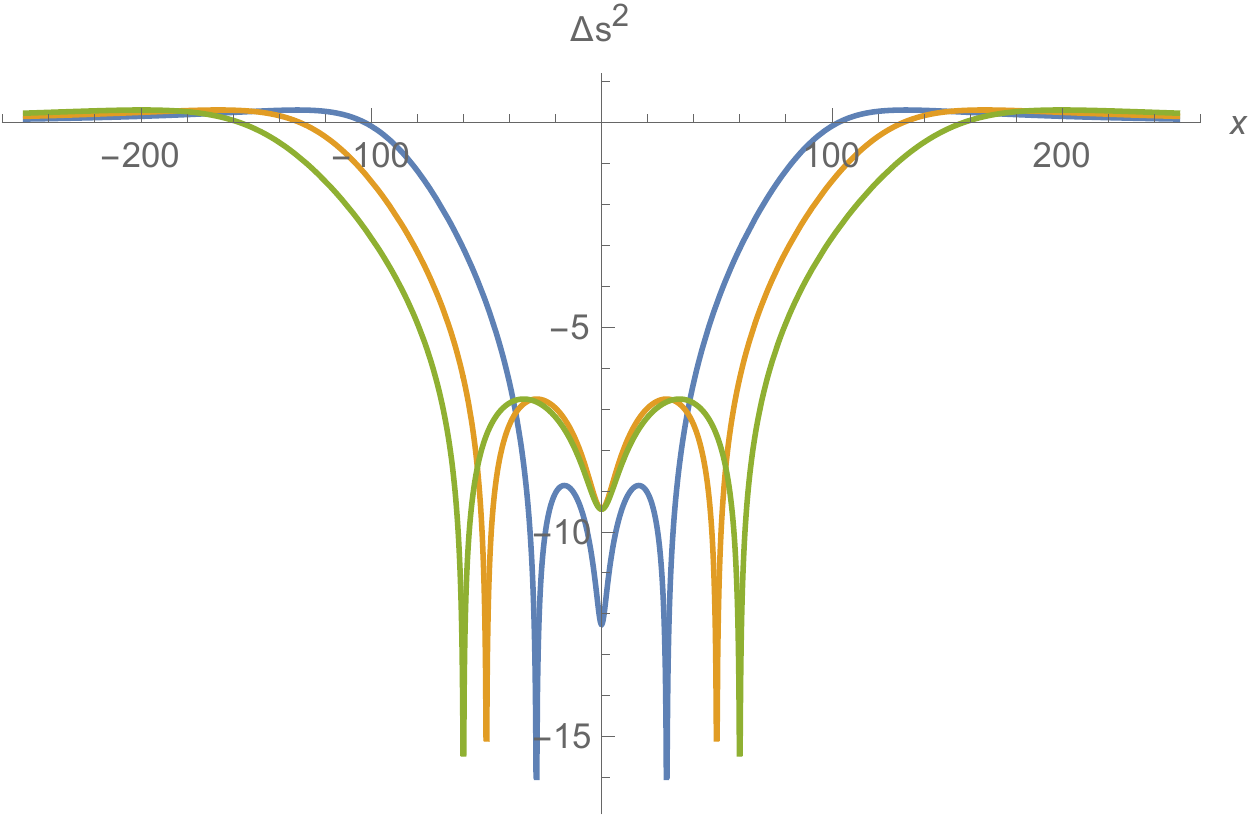}
    \caption{$\Delta S_A^{(2)}$ as a function for same components $F_{ry}$ of the field strength . Blue line : $\alpha=2$, $\alpha'=400$ ; orange line $\alpha=4$, $\alpha'=400$;  green line $\alpha=5$, $\alpha'=500$.}
    \label{plotFry}
\end{figure}

From the above plot of $\Delta S^{(2)}$, we observe that changes significantly near the boundary of the subsystem which is at $x=0$. More importantly, $\Delta S^{(2)}$ diverges at a point $x=\sqrt{\alpha\alpha'}.$ This is one of the major differences with the pseudo-entropy of the excited states created by the $F_{r\theta}$ and $F_{xz}$. Let us investigate the reason for the divergence of $\Delta S^{(2)}$ at the point $x=\sqrt{\alpha \alpha'}.$ The two-point function of $F_{ry}$ at $n=1$ sheet is given in \eqref{sh1Fry}. We can write it as a function of $x$ and Euclidean times $\alpha$ , $\alpha'$.
\begin{align}\label{Fry1}
    \langle F_{ry} F_{ry}\rangle_{\Sigma_1}&=\frac{x^2-\alpha  \alpha '}{\pi ^2 \left(\alpha -\alpha '\right)^4 \sqrt{\alpha ^2+x^2} \sqrt{\left(\alpha '\right)^2+x^2}}.
\end{align}
Note that, the two-point function vanishes at $x=\pm\sqrt{\alpha\alpha'}$. From the explicit expression of $\Delta S^{(2)}$ we see that the square of the two-point function on the same sheet comes in the denominator and therefore $\Delta S^{(2)}$ becomes singular at the point $x=\pm\sqrt{\alpha \alpha'}$. So we understand that $\Delta S^{(2)}$ decreases near the boundary which is similar to the scalar case as well as for the states excited by $F_{r\theta}$ or $F_{xz}$. The main difference turns out to be the singularity of $\Delta S^{(2)}$ at $x=\pm\sqrt{\alpha \alpha'}$ in this case. But it follows the general features of the scalar case in $d=4$ dimension except at the point $x=\pm \sqrt{\alpha \alpha'}.$ We will see the singularity at $x=\pm \sqrt{\alpha\alpha'}$ as a coordinate artifact in section \eqref{Fthy}.
\subsection*{Real-time evoulution}
 We also analyze the real-time evolution of $\Delta S^{(2)}_A$ for the same components $F_{ry}$ of field strength. We substitute $\alpha=-i t-\epsilon$ and $\alpha'=-it+\epsilon$ in the expression of $\Delta S^{(2)}_A$, where $\epsilon$ is a small positive real number.
  \begin{figure}[htp]
    \centering
    \includegraphics[width=8cm]{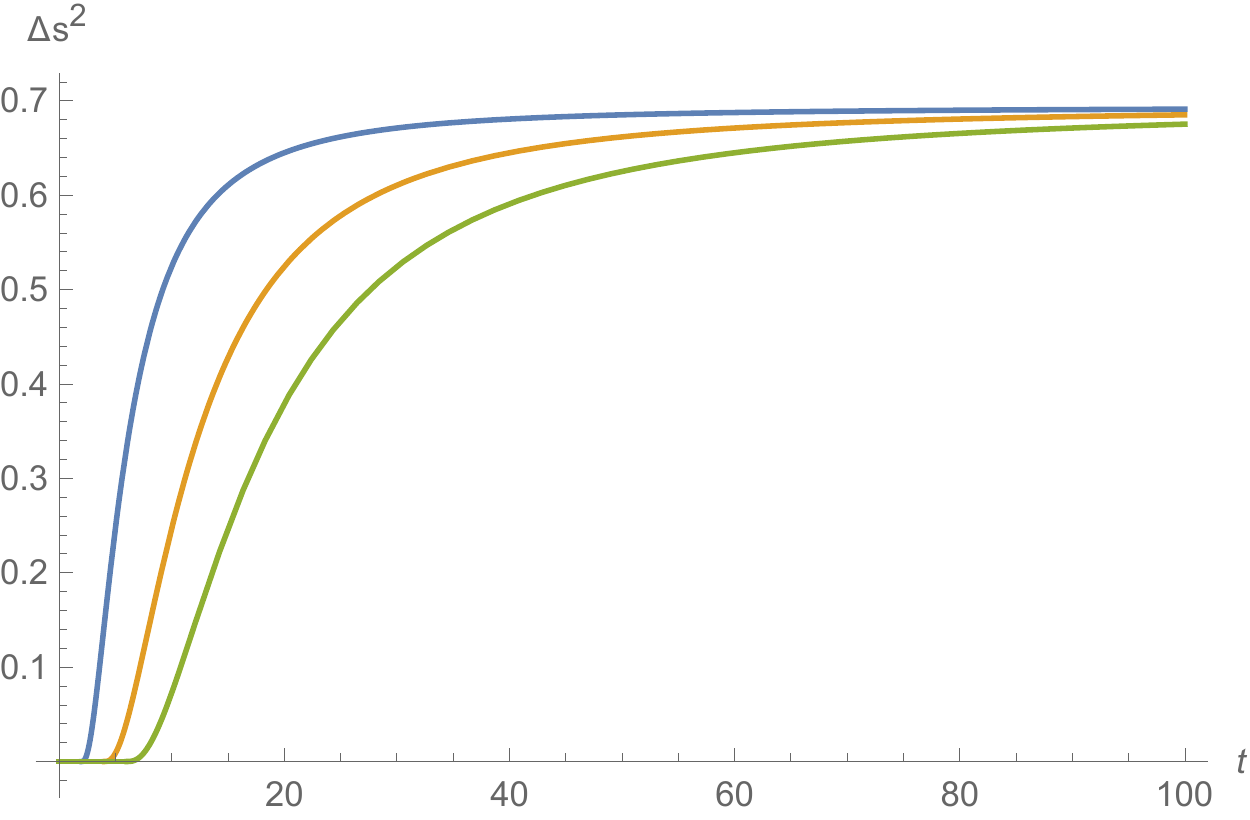}
    \caption{$\Delta S_A^{(2)}$ as a function for same components $F_{ry}$ of the field strength . Blue line : $x=2$, ; orange line $x=4$;  green line $x=6$. We keep $\epsilon=0.01$ in all cases.}
    \label{pseudos}
\end{figure}
We insert the operators at the same $y$ and $z$ co-ordinate. We fix the $x$ co-ordinate and observe the time depenence of $\Delta S^{(2)}_A$. 
We observe that in the large time $\Delta S_A^{(2)}$ reaches to $\log 2$.
\begin{align}
    \lim_{t\rightarrow \infty}\Delta S^{(2)}_A&=\log (2)-\frac{81 x^2}{16 t^2}+\cdots
\end{align}
Note that, the growth of $\Delta S^{(2)}_A$ to reach maximal entanglement due to the local quench by the operator $F_{yz}$ is different to that of $\Delta S^{(2)}_A$ by the operator $F_{r\theta}$ in the large $t$ limit and this should be because this  is a vector like excitation whereas the excitation by $F_{r\theta}$ was a pseudo-scalar excitation. However, it still saturates to $\log 2$ when left and right moving modes become maximally entangled in the large time limit.
\subsection*{Excitation created by $F_{\theta y}$}\label{Fthy}
We  consider the case where the excitations are created by two same components of the field strength at different Euclidean time $\alpha$ and $\alpha'$. In this case we choose the field strength to be $F_{\theta y}$.
 \begin{align}
    \begin{split}
        |\psi_1\rangle &=e^{-\alpha H_{\rm{CFT}}}F_{\theta y}(x_1,y_1,z_1)|0\rangle\\
        |\psi_2\rangle&=e^{-\alpha' H_{\rm{CFT}}}F_{\theta y}(x_2,y_1,z_1)|0\rangle
    \end{split}
    \end{align}
    We place the operators at the same $y$ and $z$ co-ordinates but in different $x$ coordinates. We want to study $\Delta S^{(2)}_A$  as a function of the center of the two operators. This is same as keeping the operators fixed and moving the center of the subsystem which is $x>0$ in this case. We now compute $\Delta S^{(2)}_{A}$,
 \begin{align}\label{sf}
     \Delta S^{(2)}_A&=-\log \frac{\langle F_{ \theta y}(r_1,\theta_1^{(1)},y_1,z_1)F_{ \theta y}(r_2,\theta_2^{(1)},y,z)F_{\theta y}(r_1,\theta_1^{(2)},y_1,z_1)F_{ \theta y}(r_2,\theta_2^{(2)},y_1,z_1)\rangle}{\langle F_{\theta y}(r_1,\theta_1^,y,z)F_{\theta y}(r_2,\theta_2,y_1,z_1)\rangle^2_{\Sigma_1}}
 \end{align}
 To evaluate $\Delta S^{(2)}_A$, we require the four and two-point functions of $F_{\theta y}$ on the replica surface.  Using the definition of the two-point functions given in \eqref{corm1}, we obtain
 \begin{align}\label{2ptFthy}
     \langle F_{\theta y} F_{\theta y}\rangle&=-\frac{r_1r_2}{2}\left(\partial_{r_1}\partial_{r_2}-\frac{1}{r_1r_2}\partial_{\theta_1}\partial_{\theta_2}\right)G(x_{i_1},x_{i_2})\nonumber\\
     &=-r_1r_2\langle F_{ry} F_{ry}\rangle
 \end{align}
 Note that this two-point function reflects the duality between $F_{ry}$ and $F_{\theta y}$. 
  Here $G(x_{i_1},x_{i_2})$ is the scalar two-point function on the replica surface. To evaluate $\Delta S^{(2)}_A$, we need the four-point function which is given by
  \begin{align}
    &  \langle F_{ \theta y}(r_1,\theta_1^{(1)},y_1,z_1)F_{ \theta y}(r_2,\theta_2^{(1)},y,z)F_{\theta y}(r_1,\theta_1^{(2)},y_1,z_1)F_{ \theta y}(r_2,\theta_2^{(2)},y_1,z_1)\rangle=\nonumber\\
    &\quad\qquad\qquad(r_1r_2)^2 \langle F_{ry}(r_1,\theta_1^{(1)},y,z)F_{r y}(r_2,\theta_2^{(1)},y,z)F_{ry}(r_1,\theta_1^{(2)},y,z)F_{r y}(r_2,\theta_2^{(2)},y,z)\rangle
  \end{align}
   We also evaluate the two-point function on $n=1$ sheet,
  \begin{align}
  \langle F_{\theta y}F_{\theta y}\rangle_{\Sigma_1}
  &=-\frac{r_1r_2}{2}\left(\partial_{r_1}\partial_{r_2}-\frac{1}{r_1r_2}\partial_{\theta_1}\partial_{\theta_2}\right)G(x_{i_1},x_{i_2})_{n=1}\nonumber\\
  &=-r_1r_2\langle F_{ry}F_{ry}\rangle_{\Sigma_1}
  \end{align}

From all the two-point functions, it is easy to see that $\Delta S^{(2)}_A$ for the states created by the operator $F_{\theta y}$ (inserted at two different Euclidean times) will be identical to that of $\Delta S^{(2)}_A$ for $F_{ry}$. The overall scale factor $(r_1 r_2)^2$ in the numerator gets cancelled from the denominator in \eqref{sf}. Therfore, the properties of $\Delta S^{(2)}_A$ remain the same in this case. It exhibits the similar nature shown in \eqref{plotFry}. So we observe that, at the point $x=\pm\sqrt{\alpha \alpha'}$ where two states become orthogonal to each other, $\Delta S^{(2)}_A$ diverges and it is mostly zero everywhere except near the boundary of the subsytems. To understand the orthogonality of the two states more extensively
let us also consider the case where one prepares states by acting $F_{\tau y}$ on vacuum, where $\tau$ is the Euclidean time direction. One has to evaluate the two point function of $F_{\tau y}$ on replica surface. But one can relate $F_{\tau y}$ to $F_{ry}$ and $F_{\theta y}$ by coordinate transformation.
\begin{align}
    F_{\tau y}&=\frac{\partial r}{\partial \tau}F_{ry}+\frac{\partial \theta }{\partial \tau}F_{\theta y}.
\end{align}
Therefore two-point functions of $F_{\tau y}$ can be computed from the two-point functions of $F_{ry}$ and $F_{\theta y}$. 
\begin{align}\label{2ptFty}
    \langle F_{\tau y} F_{\tau y}\rangle &=\frac{\partial r_1}{\partial \tau_1}\frac{\partial r_2}{\partial \tau_2}\langle F_{ry}F_{ry}\rangle+\frac{\partial \theta_1 }{\partial \tau_1}\frac{\partial \theta_2}{\partial \tau_2}\langle F_{\theta y}F_{\theta y}\rangle\nonumber\\
    &+\frac{\partial r_1}{\partial \tau_1}\frac{\partial \theta_2}{\partial \tau_2}\langle F_{ry}F_{\theta y}\rangle+\frac{\partial \theta_1}{\partial \tau_1}\frac{\partial r_2}{\partial \tau_2}\langle F_{\theta y}F_{r y}\rangle
\end{align}
We have computed the two-point function $\langle F_{ry}F_{ry}\rangle$ in \eqref{2ptFry} and $\langle F_{\theta y} F_{\theta y}\rangle$ in \eqref{2ptFthy}. Let us now compute $\langle F_{ry}F_{\theta y}\rangle.$
\begin{align}
  \langle F_{ry}F_{\theta y}\rangle&=\partial_{r_1}\partial_{\theta_2}\langle A_y A_y\rangle+\partial_{y_1}\partial_{y_2}\langle A_r A_{\theta}\rangle\nonumber\\
  &=\frac{1}{2}\left(\partial_{r_1}\partial_{\theta_2}+\frac{r_2}{r_1}\partial_{r_2}\partial_{\theta_1}\right)G(x_{i_1};x_{i_2})
\end{align}
Here $G(x_{i_1};x_{i_2})$ is the scalar two-point function in $d=4$ dimension on the replica surface.
Note that, we place two operators  at two different Euclidean times $\tau_1=\alpha$ and $\tau_2=\alpha'$ and therefore the two-point function of $F_{\tau y}$ becomes
\begin{align}\label{2Fty}
     \langle F_{\tau y}F_{\tau y}\rangle_{\substack{\tau_1=\alpha\\ \tau_2=\alpha'}}&=\frac{\alpha\alpha'-x^2}{2r_1r_2}\langle F_{ry}F_{ry}\rangle_{\substack{\tau_1=\alpha\\ \tau_2=\alpha'}}-\frac{(\alpha+\alpha')x}{2r_1r_2}\left(\partial_{r_1}\frac{\partial_{\theta_2}}{r_2}+\partial_{r_2}\frac{\partial_{\theta_2}}{r_1}\right)G(x_{i_1};x_{i_2})_{\substack{\tau_1=\alpha\\ \tau_2=\alpha'}}
\end{align}
To derive equation \eqref{2Fty}, we use the relation between $\langle F_{ry}F_{ry}\rangle$ and $\langle F_{\theta y}F_{\theta}\rangle$ which is given in
\eqref{2ptFthy}.
The two-point function of $F_{ry}$ on $n=1$ sheet is given in \eqref{Fry1}.
Note that, the first term vanishes at $x=\pm\sqrt{\alpha\alpha'}$. Let us compute the second term explicitly for $n=1$,
\begin{align}
   \lim_{x\rightarrow \pm\sqrt{\alpha \alpha'}}  \langle F_{ry}F_{\theta y}\rangle_{\substack{\tau_1=\alpha\\ \tau_2=\alpha'}}&=   \frac{x^2 \left(\alpha '+\alpha \right) \left(\sqrt{\alpha ^2+x^2}-\sqrt{\left(\alpha '\right)^2+x^2}\right)}{4 \pi ^2 \left(\alpha '-\alpha \right)^5 \left(\alpha ^2+x^2\right) \left(\left(\alpha '\right)^2+x^2\right)^{3/2}}\nonumber\\
  &\qquad\qquad\qquad\qquad\qquad \times  \left(\left(\alpha '+\alpha \right)^2+4 \sqrt{\left(\alpha ^2+x^2\right) \left(\left(\alpha '\right)^2+x^2\right)}+4 x^2\right)
\end{align}
We observe that the second term does not vanish at $x=\pm\sqrt{\alpha\alpha'}$ and hence the two excited states prepared by $F_{\tau y}$ acting on the vacuum will not be orthogonal at $x=\pm\sqrt{\alpha\alpha'}$. Therefore, the orthogonality of states is associated only with the components $F_{ry}$ and $F_{\theta y}$ indicating the coordinate artifact.

Now we compute $\Delta S^{(2)}_A$  for the excited states created by $F_{ry}$ or $F_{\theta y}$ acting on vacuum. We observe that near the boundary of the subsystems it depends on the ratio $p=\frac{\alpha}{\alpha'}$ of the two Euclidean times. When $p\sim 0$ or $\alpha\gg \alpha'$, $\Delta S^{(2)}_A\sim \log p^3$ and for $p\sim 1$, $\Delta S^{(2)}_A \sim -\frac{3}{128}(p-1)^4.$ 

So we understand that $\Delta S^{(2)}_A$ only changes near the boundary of the subsystems and vanishes far away from the boundary. The singularity at $x=\pm \sqrt{\alpha \alpha'}$ is just  a coordinate artifact which does not show up in other components of the field strengths.
\subsection{Excitation by the different components of the field strength with different cutoffs}
We create two different states by the different components on the field strengths acting on the ground state. We choose two different UV cutoffs. In other words the operators are placed in two different Euclidean times.

We create two states in the following way
\begin{align}
    \begin{split}
        |\psi_1\rangle&=e^{-\alpha H_{\rm{CFT}}}F_{r y}(x_1,y_1,z_1)|0\rangle\\
        |\psi_2\rangle &=e^{-\alpha' H_{\rm{CFT}}}F_{\theta y}(x_2,y_1,z_1)|0\rangle
    \end{split}
\end{align}
 We follow the same strategy and place the operators at the same $y$ and $z$ coordinates but in different $x$ coordinates. We want to study $\Delta S^{(2)}_A$  as a function of the center of the two operators. This is the same as keeping the operators fixed and moving the center of the subsystem which is $x>0$ in this case.
 We now compute $\Delta S^{(2)}_{A}$,
 \begin{align}
     \Delta S^{(2)}_A&=-\log \frac{\langle F_{r y}(r_1,\theta_1^{(1)},y_1,z_1)F_{ \theta y}(r_2,\theta_2^{(1)},y,z)F_{ry}(r_1,\theta_1^{(2)},y_1,z_1)F_{ \theta y}(r_2,\theta_2^{(2)},y_1,z_1)\rangle}{\langle F_{ry}(r_1,\theta_1^,y,z)F_{\theta y}(r_2,\theta_2,y_1,z_1)\rangle^2_{\Sigma_1}}
 \end{align}
 To evaluate $\Delta S^{(2)}_A$, we require the four and two-point functions of $F_{ry}$ and $F_{\theta y}$ on the replica surface.  Using the definition of the two-point functions given in \eqref{corm1}, we obtain
 \begin{align}
     \langle F_{ry} F_{\theta y}\rangle&=\left(\partial_{r_1}\partial_{\theta_2}\langle A_y A_y\rangle +\partial_{y_1}\partial_{y_2}\langle A_r A_{\theta}\rangle\right)\nonumber\\
     &=\frac{1}{2}\left(\partial_{r_1}\partial_{\theta_2}+\frac{r_2}{r_1}\partial_{r_2}\partial_{\theta_1}\right)G(x_{i_1},x_{i_2})
 \end{align}
 To derive the last line we use the isotropic condition in the $y-z$ plane given in \eqref{isot}.
 Similarly we also need the two-point functions of $F_{ry}$ and $F_{\theta y}$. The two-point functions of $F_{r y}$ is  given in \eqref{2ptFry}. Two-point function of $F_{\theta y}$ is also given in \eqref{2ptFthy}. 
 We also evaluate the two-point function on $n=1$ sheet,
  \begin{align}
  \langle F_{ry}F_{\theta y}\rangle_{\Sigma_1}&=\frac{1}{2}\left(\partial_{r_1}\partial_{\theta_2}+\frac{r_2}{r_1}\partial_{r_2}\partial_{\theta_1}\right)G(x_{i_1},x_{i_2})_{n=1}\nonumber\\
  &=\frac{r_2 \left(r_2-r_1\right) \left(r_1+r_2\right) \sin \left(\theta _1-\theta _2\right)}{\pi ^2 \left(-2 r_2 r_1 \cos \left(\theta _1-\theta _2\right)+r_1^2+r_2^2\right){}^3}
  \end{align}
  To evaluate $\Delta S^{(2)}_A$, we compute the four-point function
  \begin{align}
        &  \langle F_{ry}(r_1,\theta_1^{(1)},y,z)F_{\theta y}(r_2,\theta_2^{(1)},y,z)F_{ry}(r_1,\theta_1^{(2)},y,z)F_{\theta y}(r_2,\theta_2^{(2)},y,z)\rangle=\nonumber\\
     & \qquad\qquad\qquad\qquad \langle F_{ry}(r_1,\theta_1^{(1)},y,z)F_{\theta y}(r_2,\theta_2^{(1)},y,z)\rangle\langle F_{ry}(r_1,\theta_1^{(2)},y,z)F_{\theta y}(r_2,\theta_2^{(2)},y,z)\rangle+\nonumber\\
     &\qquad\qquad\qquad\qquad\langle  F_{ry}(r_1,\theta_1^{(1)},y,z)F_{ry}(r_1,\theta_1^{(2)},y,z)\rangle \langle F_{\theta y}(r_2,\theta_2^{(1)},y,z)F_{\theta y}(r_2,\theta_2^{(2)},y,z) \rangle +\nonumber\\
     &\qquad\qquad\qquad\qquad \langle F_{ry}(r_1,\theta_1^{(1)},y,z)F_{\theta y}(r_2,\theta_2^{(2)},y,z)\rangle\langle F_{\theta y}(r_2,\theta_2^{(1)},y,z)F_{r y}(r_1,\theta_1^{(2)},y,z)\rangle \nonumber\\
     &=\frac{1}{4}\left(\partial_{r_1}\partial_{\theta_2}+\frac{r_2}{r_1}\partial_{r_2}\partial_{\theta_1}\right)G(r_1,\theta_1;r_2,\theta_2)\left(\partial_{r_1}\partial_{\theta_2}+\frac{r_2}{r_1}\partial_{r_2}\partial_{\theta_1}\right)G(r_1,\theta_1;r_2,\theta_2)\nonumber\\
     &-\lim_{\substack{r_1\rightarrow r_2\\ \theta_1\rightarrow\theta_2+2\pi}}\frac{r_1r_2}{4}\left(\partial_{r_1}\partial_{r_2}-\frac{1}{r_1r_2}\partial_{\theta_1}\partial_{\theta_2}\right)G(r_1,\theta_1;r_2,\theta_2)\left(\partial_{r_1}\partial_{r_2}-\frac{1}{r_1r_2}\partial_{\theta_1}\partial_{\theta_2}\right)G(r_1,\theta_1;r_2,\theta_2)\nonumber\\
     &+\frac{1}{4}\left(\partial_{r_1}\partial_{\theta_2}+\frac{r_2}{r_1}\partial_{r_2}\partial_{\theta_1}\right)G(r_1,\theta_1;r_2,\theta_2+2\pi)\left(\partial_{r_1}\partial_{\theta_2}+\frac{r_2}{r_1}\partial_{r_2}\partial_{\theta_1}\right)G(r_1,\theta_1;r_2,\theta_2+2\pi)\nonumber\\
     &=\frac{\left(r_1-r_2\right){}^2 r_2 \sin ^2\left(\frac{1}{2} \left(\theta _1-\theta _2\right)\right) \left(2 \sqrt{r_1 r_2} \cos \left(\frac{1}{2} \left(\theta _1-\theta _2\right)\right)-3 \left(r_1+r_2\right)\right){}^2}{256 \pi ^4 r_1 \left(r_1+r_2\right){}^4 \left(-2 \sqrt{r_1 r_2} \cos \left(\frac{1}{2} \left(\theta _1-\theta _2\right)\right)+r_1+r_2\right){}^6}\nonumber\\
     &-(\frac{3}{256\pi^2})^2\frac{1}{r_1^4r_2^2}\nonumber\\
     &+\frac{\left(r_1-r_2\right){}^2 \sin ^2\left(\frac{1}{2} \left(\theta _1-\theta _2\right)\right) \left(2 \sqrt{r_1 r_2} \cos \left(\frac{1}{2} \left(\theta _1-\theta _2\right)\right)+3 \left(r_1+r_2\right)\right){}^2}{256 \pi ^4 \left(r_1+r_2\right){}^4 \left(2 \sqrt{r_1 r_2} \cos \left(\frac{1}{2} \left(\theta _1-\theta _2\right)\right)+r_1+r_2\right){}^6}
  \end{align}
  Here, $G(r_1,\theta_1;r_2,\theta_2)$ is the scalar two-point function on the replica surface for $n=2$.
  Note that, the four point function diverges negatively as the operator approaches to the boundary which is located at $x=0$. This is due to the second term which comes from the correlators across the sheets but involve same points. Therefore, $\Delta S^{(2)}_A$ becomes complex near the boundary. This is one of the examples where $\Delta S^{(2)}_A$ becomes complex and the reason is that the reduced transition matrix is not Hermitian which can be seen from the expression
\begin{align}
   \ra&=\rm{Tr}_B  \left(F_{ry}(\alpha,\textbf{x})|0\rangle\langle 0|F_{\theta y}(\alpha',\textbf{x})\right). 
\end{align}
 Now we plot the real part of $\Delta S^{(2)}_A$ as a function of the center of the two operators.
 \begin{figure}[htp]
    \centering
    \includegraphics[width=8cm]{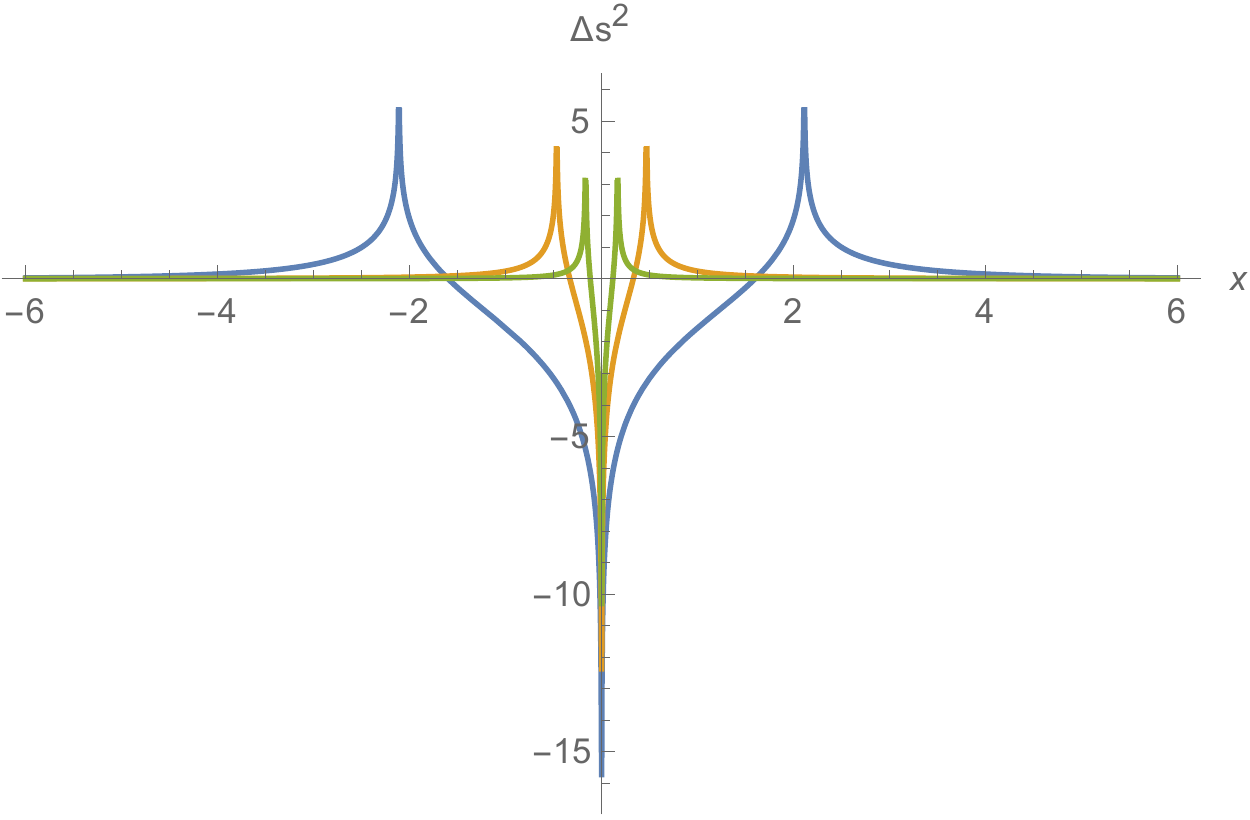}
    \caption {Real part of $\Delta S_A^{(2)}$ as a function for different components of the field strength . Blue line : $\alpha=5$, $\alpha'=25$ ; orange line $\alpha=10$, $\alpha'=30$;  green line $\alpha=15$, $\alpha'=35$.}
    \label{plotFrythy}
\end{figure}

From figure \eqref{plotFrythy}, we observe that real part of $\Delta S^{(2)}_A$ decreases significantly near the boundary of the subsystems and vanishes far away from the boundary. Therefore, there is no contribution to $\Delta S^{(2)}_A$ far away from the boundary where pseudo \Re entropy becomes equal to the \Re entropy of the ground state.

\section{Discussion}
In this paper, we study pseudo entropy in the free Maxwell field theory in 
$d=4$ dimension. Mainly we are interested in evaluating the difference between the pseudo \Re entropy and the \Re entropy of the ground states. This effectively captures the variation of the \Re entropy from the ground state.

To set up the whole formalism we begin with conformal scalar field theory in $d=4$ dimension. We prepare two excited states by two conformal operators with fixed conformal weights acting on the ground state. We keep the spatial positions separate them by placing them at two different Euclidean times. We observe that the difference between the pseudo entropy and the ground state \Re entropy are the same everywhere except near the boundary of the subsystems where it changes significantly. This difference at the boundary actually depends on the ratio of the two Euclidean times near the boundary. Near boundary, behavior can be understood at the correlator level. The two-point functions on the replica surface change significantly near the boundary and hence it is reflected on the $\Delta S^{(n)}_A.$ We also show that under a suitable analytical continuation of pseudo \Re entropy leads to evaluation of real-time evolution of \Re entropy during quenches. In this case $\Delta S^{(n)}_A$ starts growing from the point when real time of the operator becomes the same as the spatial insertion point and it reaches to $\log 2$ after a large time when left and right moving modes become maximally entangled \cite{Nozaki:2014hna}.

To understand the general features of $\Delta S^{(n)}_A$ in gauge theory, we prepare excited states by a different component of the field strengths acting on the vacuum. Therefore the states remain gauge invariant and we evaluate the difference between pseudo \Re entropy and the \Re entropy of the ground state. Similar to the scalar field, the difference $\Delta S^{(n)}_A$ is mostly zero everywhere except near the boundary of the subsystems. This property can be explained by the two-point functions of the field strength on the replica surface. Two-point correlators also exhibit a significant change near the boundary which reflects on $\Delta S^{(n)}_A$ and the peak of $\Delta S^{(n)}_A$ depends on the ratio of the Euclidean times of the field strengths.

As a future direction, we would like to investigate pseudo entropy for the linearized graviton. One can, in principle, create excitations using the Riemann tensor acting on the ground state. The difference between pseudo entropy from the ground state entropy can be evaluated in Euclidean path integral formalism. Therefore one can also compare $\Delta S^{(n)}_A$ for spin-$0$, spin-$1$ and spin-$2$ field and understand the general spin dependence. The physical question would be to relate this quantity $\Delta S^{(n)}_A$ with some property of the local operator which creates the excitation. It will be interesting to evaluate and understand the general properties of pseudo-entropy for the fermionic systems where one can prepare different excited states by the different primaries at different Euclidean times. But Wick rotating the Euclidean time to real-time should lead to the analysis of the local quench by the fermionic operators \cite{Nozaki:2015mca}. Another important direction would be to understand pseudo-entropy in conformal higher derivative and conformal higher spin fields \cite{Mukherjee:2021alj,Mukherjee:2021rri} where one has to develop two and four-point functions on the replica surface. It will be nice to show the non-unitarity nature of these theories within the framework of pseudo entropy.
 \acknowledgments
It is a pleasure to thank Justin R. David, Aninda Sinha, and Chethan Krishnan for fruitful
discussions and comments on the draft.
    \appendix
    \section{Derivation of two-point functions and  $\Delta S^{(n)}_A$ for arbitary $n$}\label{appA}
   In this section we present the derivation for $\Delta S^{(n)}_A$, for scalar field, free Maxwell field in $d=4$ dimension. The key ingredient is just the two-point functions on the replica surface. We list out the two-point functions of scalar and different components of the field strength on the replica surface for arbitrary $n$.
   \subsection*{Scalar correlator in $d=4$ dimension}
   One has to compute the $2n$-point function of conformal scalars on replica surface. Since the theory is free, one can easily evaluate it using Wick contraction of the two-point function. The two-point function on the replica surface is known \cite{Nozaki:2014hna,Nozaki:2014uaa} and  presented in \eqref{sclg}. This can also be written as
   \begin{align}\label{2ptg}
       G(x_{i_1};x_{i_2})_{(n,k)}&=\frac{\sinh\frac{\eta}{n}}{8\pi^2n r_1r_2\sinh\eta\left(\cosh\frac{\eta}{n}-\cos\frac{\theta_1-\theta_2-2\pi k}{n}\right)}
   \end{align}
   where $\cosh\eta=\frac{r_1^2+r_2^2+(y_1-y_2)^2+(z_1-z_2)^2}{2r_1r_2}.$ Note that, the two-point function which involves points separated by $k$ sheets are the same as the two-point functions separated by $k-n$ sheets.
   $$G(x_{i_1};x_{i_2})_{(n,k)}=G(x_{i_1};x_{i_2})_{(n,k-n)}.$$In this two-point function, $k=0$ denotes the correlator involving the points on the same sheet. Threrefore, $2n$-point function on replica surface can be expressed as
   \begin{align}
       \langle \phi^{(1)}(x_{i_1})\phi^{(1)}(x_{i_2})\cdots \phi^{(n)}(x_{i_1})\phi^{(n)}(x_{i_2})\rangle_{\Sigma_n}&= \left( G(x_{i_1};x_{i_2})_{(n,k=0)}\right)^n+\left( G(x_{i_1};x_{i_2})_{(n,k=1)}\right)^n+\cdots
   \end{align}
   where $\cdots$ includes all possible contraction of points in sheets for $k\geq1$. One can easily obtain the $2n$-point function from the correlator given in \eqref{2ptg}.
   \subsection*{Two-point functions of field strength on replica surface}
   We begin with the correlator $\langle F_{r\theta} F_{r\theta}\rangle$ which is proportional to the scalar Laplacian acting on the two-point function of conformal scalar.
   \begin{align}
       \langle F_{r\theta} F_{r\theta}\rangle_{\Sigma_n}&=-r_1r_2\left(\partial_{r_1}^2+\frac{1}{r_1}\partial_{r_1}+\frac{1}{r_1^2}\partial_{\theta_1}^2\right)G(x_{i_1},x_{i_2})_{(n,k)}\nonumber\\
       &=-\frac{(\coth (\eta )+1) \text{csch}(\eta ) \left(\cosh \left(\frac{\eta }{n}\right) \cos \left(\frac{\theta }{n}\right)+n \coth (\eta ) \sinh \left(\frac{\eta }{n}\right) \left(\cosh \left(\frac{\eta }{n}\right)-\cos \left(\frac{\theta }{n}\right)\right)-1\right)}{4 \pi ^2 n^2 r_1^2 \left(\cos \left(\frac{\theta }{n}\right)-\cosh \left(\frac{\eta }{n}\right)\right)^2}
   \end{align}
   Here $\theta=\theta_1-\theta_2-2\pi k.$ Using this two-point function, one can compute $2n$-point function on the replica surface. Similarly the two-point function $\langle F_{yz} F_{yz}\rangle$ on replica surface is given by
   \begin{align}
       \langle F_{yz} F_{yz}\rangle_{\Sigma_n} &=\frac{1}{2}\left(\partial_{r_1}^2+\frac{1}{r_1}\partial_{r_1}+\frac{1}{r_1^2}\partial_{\theta_1}^2\right)G(x_{i_1},x_{i_2})_{(n,k)}\nonumber\\
       &=\frac{(\coth (\eta )+1) \text{csch}(\eta ) \left(\cosh \left(\frac{\eta }{n}\right) \cos \left(\frac{\theta }{n}\right)+n \coth (\eta ) \sinh \left(\frac{\eta }{n}\right) \left(\cosh \left(\frac{\eta }{n}\right)-\cos \left(\frac{\theta }{n}\right)\right)-1\right)}{8 \pi ^2 n^2 r_1^3 r_2\left(\cos \left(\frac{\theta }{n}\right)-\cosh \left(\frac{\eta }{n}\right)\right)^2}
   \end{align}
   with $\theta=\theta_1-\theta_2-2\pi k$.
   \begin{align}
       \langle F_{ry} F_{ry}\rangle_{\Sigma_n}&=\frac{\text{csch}(\eta )}{16 \pi ^2 n^3 r_1^2 r_2^2 \left(\cos \left(\frac{\theta }{n}\right)-\cosh \left(\frac{\eta }{n}\right)\right)^3}\Big[8 n \coth (\eta ) \cosh ^2\left(\frac{\eta }{n}\right) \cos \left(\frac{\theta }{n}\right)\nonumber\\
      & -2 n \cosh \left(\frac{\eta }{n}\right)-4 n \coth (\eta ) \cosh ^3\left(\frac{\eta }{n}\right)\coth (\eta ) \left(-\cosh \left(\frac{2 \eta }{n}\right)+\cos \left(\frac{2 \theta }{n}\right)+2\right)\nonumber\\
      &+\sinh \left(\frac{\eta }{n}\right)\big(-3-3 \coth ^2(\eta )+\big(\left(2 n^2-1\right) \cos \left(\frac{2 \theta }{n}\right)+4 n^2+3\big)\nonumber\\
      &+2 n^2 \cosh \left(\frac{2 \eta }{n}\right)+\cosh (2 \eta ) +\cos \left(\frac{2 \theta }{n}\right)\text{csch}^2(\eta )-4 n \coth (\eta ) \sinh \left(\frac{\eta }{n}\right) \cos \left(\frac{\theta }{n}\right)\big)\nonumber\\
      &+\text{csch}^2(\eta ) \sinh \left(\frac{2 \eta }{n}\right) \left(\cosh (2 \eta )-4 n^2-1\right) +\cos \left(\frac{\theta }{n}\right)\Big]\nonumber\\
      &=-r_1r_2\langle F_{\theta y}F_{\theta y}\rangle_{\Sigma_n}.
   \end{align}
   Given all the two-point functions on the replica surface for arbitrary $n$, one can evaluate the $2n$-point function explicitly and compute $\Delta S^{(n)}_A.$
\bibliographystyle{JHEP}
\bibliography{pseudo_entropy} 
\end{document}